\documentclass[sigconf, nonacm]{acmart}

\newcommand\vldbdoi{XX.XX/XXX.XX}
\newcommand\vldbpages{XXX-XXX}
\newcommand\vldbvolume{14}
\newcommand\vldbissue{1}
\newcommand\vldbyear{2020}
\newcommand\vldbavailabilityurl{URL_TO_YOUR_ARTIFACTS}
\newcommand\vldbpagestyle{plain} 

\usepackage{booktabs}   %% For formal tables:
\usepackage{subcaption} %% For complex figures with subfigures/subcaptions
\usepackage[vlined,linesnumbered,ruled]{algorithm2e}
\usepackage{algpseudocode} 
\usepackage{amsmath,amsfonts,amsthm}
\usepackage{xspace}
\usepackage{multirow}
\usepackage{bbding}
\usepackage{wrapfig}
\usepackage{array}
\usepackage{xparse}
\usepackage{bbold}
\usepackage{color}
\usepackage{booktabs}
\usepackage{subcaption}
\usepackage{listings}
\usepackage{xcolor}
\definecolor{commentgreen}{RGB}{2,112,10}
\definecolor{eminence}{RGB}{108,48,130}
\definecolor{weborange}{RGB}{255,165,0}
\definecolor{frenchplum}{RGB}{129,20,83}

\newcommand{\stitle}[1]{\vspace{0.5ex}\noindent{\bf #1}}

\newcommand{\kw}[1]{\textsf{#1}\xspace}

\newcommand{\eat}[1]{}

\newcommand{\sstab}{\rule{0pt}{8pt}\\[-1.8ex]}

\newcommand{\revise}[1]{{#1}}

\newcommand{\gar}{\kw{GraphAr}}
\newcommand{\parquet}{Parquet}

\newsavebox{\bigimage}

\definecolor{codegreen}{rgb}{0,0.6,0}
\definecolor{codegray}{rgb}{0.5,0.5,0.5}
\definecolor{codered}{rgb}{0.59,0.0,0.09}
\definecolor{backcolour}{rgb}{0.96,0.96,0.96}
\definecolor{codeorange}{rgb}{1.0,0.2,0.0}
\definecolor{commentgreen}{RGB}{2,112,10}
\makeatletter
\def\UrlAlphabet{%
    \do\a\do\b\do\c\do\d\do\e\do\f\do\g\do\h\do\i\do\j%
    \do\k\do\l\do\m\do\n\do\o\do\p\do\q\do\r\do\s\do\t%
    \do\u\do\v\do\w\do\x\do\y\do\z\do\A\do\B\do\C\do\D%
    \do\E\do\F\do\G\do\H\do\I\do\J\do\K\do\L\do\M\do\N%
    \do\O\do\P\do\Q\do\R\do\S\do\T\do\U\do\V\do\W\do\X%
    \do\Y\do\Z}
\def\UrlDigits{\do\1\do\2\do\3\do\4\do\5\do\6\do\7\do\8\do\9\do\0}
\g@addto@macro{\UrlBreaks}{\UrlOrds}
\g@addto@macro{\UrlBreaks}{\UrlAlphabet}
\g@addto@macro{\UrlBreaks}{\UrlDigits}
\makeatother

\newcommand{\revisenote}[1]{}

\newtheorem{theorem}{Theorem}

\theoremstyle{definition}
\newtheorem{definition}{Definition}

\theoremstyle{remark}

\theoremstyle{property}

\begin{document}
\title{\gar: An Efficient Storage Scheme for Graph Data in\\ Data Lakes}

\author{Xue Li}
\affiliation{%
  \institution{Alibaba Group}
}

\author{Weibin Zeng}
\affiliation{%
  \institution{Alibaba Group}
}

\author{Zhibin Wang}
\affiliation{%
  \institution{Nanjing University}
}

\author{Diwen Zhu}
\affiliation{%
  \institution{Alibaba Group}
}

\author{Jingbo Xu}
\affiliation{%
  \institution{Alibaba Group}
}

\author{Wenyuan Yu}
\affiliation{%
  \institution{Alibaba Group}
}

\author{Jingren Zhou}
\affiliation{%
  \institution{Alibaba Group}
}
\email{graphar@alibaba-inc.com}

%%
%% The "author" command and its associated commands are used to define the authors and their affiliations.
\eat{
  \author{Xue Li$^1$, Weibin Zeng$^1$, Zhibin Wang$^2$, Diwen Zhu$^1$, Jingbo Xu$^1$, Wenyuan Yu$^1$, Jingren Zhou$^1$}
  \affiliation{%
    \institution{\textsuperscript{1}Alibaba Group, \textsuperscript{2}Nanjing University}
    %\streetaddress{P.O. Box 1212}
    %\state{Ireland}
  }
  \email{graphar@alibaba-inc.com}
}

\begin{abstract}
  Data lakes, increasingly adopted for their ability to store and analyze diverse types of data,
  commonly use columnar storage formats like Parquet and ORC for handling relational tables.
  However, these traditional setups fall short when it comes to efficiently managing
  graph data, particularly those conforming to the Labeled Property Graph (LPG) model.
  To address this gap, this paper introduces \gar, a specialized storage scheme designed
  to enhance existing data lakes for efficient graph data management. Leveraging the strengths
  of Parquet, \gar~captures LPG semantics precisely and facilitates graph-specific operations
  such as neighbor retrieval and label filtering. Through innovative data organization, encoding,
  and decoding techniques, \gar~dramatically improves performance. 
  %Our evaluations reveal that \gar~outperforms conventional Parquet and Acero-based methods, achieving an average speedup of $3283\times$ for neighbor retrieval, $6.0\times$ for label filtering, and $29.5\times$ for end-to-end workloads. 
  Our evaluations reveal that \gar~outperforms conventional Parquet and Acero-based methods, achieving an average speedup of $4452\times$ for neighbor retrieval, $14.8\times$ for label filtering, and $29.5\times$ for end-to-end workloads. 
  These findings highlight \gar's potential to extend the utility of
  data lakes by enabling efficient graph data management.
\end{abstract}

\maketitle

\pagestyle{\vldbpagestyle}
\begingroup\small\noindent\raggedright\textbf{PVLDB Reference Format:}\\
%\vldbauthors. 
Xue Li, Weibin Zeng, Zhibin Wang, Diwen Zhu, Jingbo Xu, Wenyuan Yu, Jingren Zhou.
\gar: An Efficient Storage Scheme for Graph Data in Data Lakes. PVLDB, \vldbvolume(\vldbissue): \vldbpages, \vldbyear.\\
\href{https://doi.org/\vldbdoi}{doi:\vldbdoi}
\endgroup
\begingroup
\renewcommand\thefootnote{}\footnote{\noindent
  This work is licensed under the Creative Commons BY-NC-ND 4.0 International License. Visit \url{https://creativecommons.org/licenses/by-nc-nd/4.0/} to view a copy of this license. For any use beyond those covered by this license, obtain permission by emailing \href{mailto:info@vldb.org}{info@vldb.org}. Copyright is held by the owner/author(s). Publication rights licensed to the VLDB Endowment. \\
  \raggedright Proceedings of the VLDB Endowment, Vol. \vldbvolume, No. \vldbissue\ %
  ISSN 2150-8097. \\
  \href{https://doi.org/\vldbdoi}{doi:\vldbdoi} \\
}\addtocounter{footnote}{-1}\endgroup

\ifdefempty{\vldbavailabilityurl}{}{
  \vspace{.3cm}
  \begingroup\small\noindent\raggedright\textbf{PVLDB Artifact Availability:}\\
  %The source code, data, and/or other artifacts have been made available at \url{\vldbavailabilityurl}.
  The source code, data, and/or other artifacts have been made available at \url{https://github.com/apache/incubator-graphar/tree/research}.
  \endgroup
}
\section{Introduction}
\label{sec:intro}

\eat{
   1. why Data Lake? why LPG? why together?\\
   - 1.1. data lake is a popular trend.\\
   - 1.2. graph data is important, LPG is a standard.\\
   - 1.3. a large amount of graph data exists in data lakes.\\
   2. using a real and specific example to demonstrate the importance of querying LPG within data lakes.\\
   3. problems in existing solutions.\\
   - 3.1. existing tabular formats failed to represent LPG semantics.\\
   - 3.2. existing tabular formats leads to poor performance when querying graphs. \\
   4. our solution (contributions of this paper.) \\
   - 4.1. using novel data organization, metadata management and utilizing existing formats as payload formats.\\
   - 4.2. identifying two critical operations in graph queries and proposing efficient solutions.\\
}
Data lakes have quickly become an essential infrastructure for organizations looking to store and analyze diverse datasets in their raw formats~\cite{Zaharia0XA21, DeltaLake, Snowflake, data_lake_management, miloslavskaya2016big, khine2018data, ramakrishnan2017azure}. As centralized repositories, they offer unparalleled flexibility in accommodating a wide array of data types, from structured relational tables to unstructured logs and text files. Crucially, they serve as a cost-effective solution for archiving data at scale while still allowing for queries on archived or rarely accessed data. This dual utility makes them invaluable for both real-time analytics and long-term data management. Columnar storage formats like Parquet~\cite{parquet} and ORC~\cite{orc} have become standard for storing tabular data in data lakes due to their robust compression and efficient query capabilities.

In sync with these trends, graph data has become increasingly importance, especially for modeling complex relationships. Leading graph-related systems like Neo4j~\cite{neo4j}, TigerGraph~\cite{TigerGraph}, JanusGraph~\cite{janusgraph} and GraphScope~\cite{graphscope} leverage the Labeled Property Graph (LPG) model~\cite{angles2023pg, baken2020linked, anikin2019labeled} for this purpose. The recent ISO SQL:2023 standard includes a SQL/PGQ extension that not only facilitates querying LPGs but also enables the creation of LPG views from relational tables~\cite{SQL2023}. This groundbreaking inclusion highlights the growing convergence of relational and graph data models and emphasizes the need to integrate LPGs into data lakes. Consequently, LPGs are making their way into data lakes for multiple uses, from backups and archives for existing graph databases to natural extensions of transactional, log, and tabular data.

Managing and analyzing LPGs in data lakes offers several significant benefits.
Graph-specific queries are often more naturally articulated in languages like Cypher~\cite{cypher_paper}, Gremlin~\cite{gremlin}, GQL~\cite{GQL}, or SQL/PGQ, providing an intuitive framework for conducting comprehensive analysis of entity relationships, facilitating the discovery of valuable insights.
Data lakes also provide computational flexibility for running complex graph algorithms, enabling the exploration of intricate patterns. Additionally, they offer cost-effective storage solutions, allowing organizations to utilize more affordable and colder storage options without sacrificing query performance. Most notably, data lakes enable seamless querying across both graph and relational data, ushering in a holistic approach to data analytics.

\begin{figure}[tbp]
   \centering
   \includegraphics[width=\linewidth]{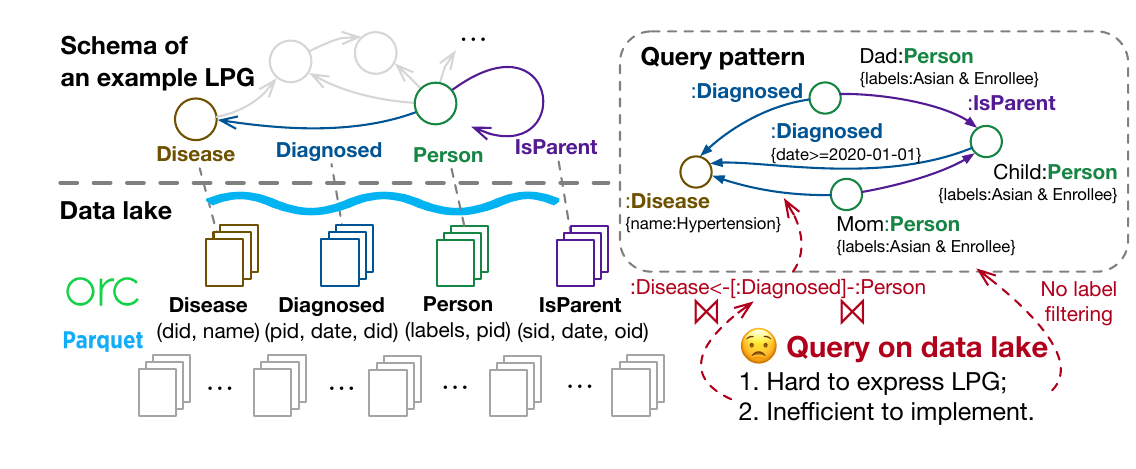}
   \vspace{-9mm}
   \caption{A graph-related query within the data lake.}
   \vspace{-3mm}
   \label{fig:example}
\end{figure}

As shown in Figure~\ref{fig:example}, the example workload illustrates a scenario of immense relevance to public health researchers. The query aims to count the number of families—comprising a father, mother, and child—each labeled as Asian and Enrollee (indicating their participation in a health study), and diagnosed with hypertension since 2020. Such queries hold significant utility for public health studies as they allow for the analysis of correlations between familial relationships, racial groups, and specific health conditions like hypertension among study participants. Understanding these relationships can be critical for targeted health interventions and for identifying possible social or genetic factors contributing to disease prevalence. Within the context of this research query, data lakes offer an economical and scalable solution for storing diverse, multi-source, and often historical health-related data. More importantly, the intricate relationships and multiple attributes required by this research are more naturally and efficiently captured through property graph queries than through traditional SQL.
However, integrating LPGs into data lakes introduces unique challenges:

\revisenote{M2 R2.W1 R2.D1 R3.W1}
\revise{
\sstab
%Firstly, there is no standardized way to encapsulate an LPG within the existing data lake architecture. While columnar formats like Parquet and ORC excel at storing individual tables, they fall short in representing the complex relationships and semantics across these tables, which are inherent to LPGs.
\textbf{Challenge 1:} There is no standardized way to encapsulate an LPG within the existing data lake architecture. While columnar formats like Parquet and ORC excel at storing individual tables, they fall short in representing the complex relationships and semantics across these tables, which are inherent to LPGs.

\sstab
\textbf{Challenge 2:} Graph-specific operations can be highly inefficient in this setup. The foundational operation, neighbor retrieval, might require multiple joins, significantly impacting performance.

\sstab
\textbf{Challenge 3:} Label filtering, another essential graph-specific operation, also introduces inefficiency due to the lack of native support in columnar formats.

%\sstab
%Secondly, executing graph-specific operations, particularly neighbor retrieval and label filtering, can be highly inefficient in this setup. For example, neighbor retrieval might require multiple joins, significantly impacting performance. Label filtering, another essential operation, also introduces inefficiency due to the lack of native support in columnar formats.
}

\eat{
These challenges underline the pressing need for specialized storage formats and analytical tools that can seamlessly integrate LPGs into existing data lake infrastructures. 
Parquet and ORC provide flexible and efficient support for various datatypes, including atomic types (e.g., bools and integers), and nested and/or repeated structures (e.g., arrays and maps), as well as encoding techniques like delta encoding, run-length encoding, etc.
Previous studies have demonstrated the importance of leveraging these capabilities to optimize data management~\cite{floratou2014sql, ivanov2020impact, DeltaLake}.
Besides, recent research~\cite{li2023selection, Vectorizing2013, simd4, FlexPushdownDB, PushdownDB, AWS_S3_Select} has highlighted the significant benefits of instruction sets like BMI and SIMD for storage predicate pushdown and execution acceleration on Intel and AMD CPUs. 
This motivates us to leverage these instructions to enhance the performance of graph-specific operations.
Thus, a promising solution to address the above challenges would be to utilize the efficiencies of columnar formats like Parquet and ORC while addressing the performance bottlenecks inherent in current systems.

This paper introduces \gar~(\textbf{E}fficient stor\textbf{A}ge for \textbf{G}raph data in data \textbf{L}ak\textbf{E}s), a novel storage scheme designed to enhance data lakes' capabilities in managing LPGs efficiently. The paper's primary contributions are:
}

\stitle{\gar}. To address these challenges, we introduce \gar\footnote{Apache GraphAr is an effort undergoing incubation at the Apache Software Foundation (ASF), sponsored by the Apache Incubator.}, an efficient storage scheme for graph data in data lakes.
It is designed to enhance the efficiency of data lakes utilizing the capabilities of existing formats, with a specific focus on Parquet in this paper. 
\gar~ensures seamless integration with existing tools and introduces innovative additions specifically tailored to handle LPGs.

\revise{
\sstab
Firstly, Parquet provides flexible and efficient support for various datatypes, including atomic types (e.g., bools and integers), and nested and/or repeated structures (e.g., arrays and maps).
Leveraging Parquet as its fundamental building block, \gar~further introduces standardized YAML files to represent the schema metadata for LPGs, alongside a hierarchical data layout to store the data. 
%Additionally, \gar incorporates optimized layouts that accommodate the access patterns commonly associated with graph data. 
This innovative combination of data organization with metadata management enables the complete expression of LPG semantics, while ensuring compatibility with both data lake ecosystems and existing graph-related systems, addressing \textbf{Challenge 1}.

\sstab
%Secondly, \gar incorporates specialized optimization techniques to improve the performance of two critical graph operations: neighbor retrieval and label filtering, which are not inherently optimized in existing formats.
\gar incorporates specialized optimization techniques to improve the performance of critical graph operations, which are not inherently optimized in existing formats.
To address \textbf{Challenge 2}, \gar facilitates neighbor retrieval by organizing edges as sorted tables in Parquet to enable an efficient CSR (Compressed Sparse Row) or CSC (Compressed Sparse Column)-like representation, and leveraging Parquet's delta encoding to reduce overhead in data storage and loading.
\gar also introduces an innovative decoding algorithm that utilizes instruction sets such as BMI (Bit Manipulation Instructions) and SIMD (Single Instruction, Multiple Data), along with a unique structure named PAC (Page-Aligned Collections), to further accelerate the neighbor retrieval process.

\sstab
In addressing \textbf{Challenge 3} of label filtering, \gar adapts the RLE (Run-Length Encoding) technique from Parquet and introduces a novel interval-based decoding algorithm. 
Through integrating proven methods (CSR/CSC, delta encoding, RLE) with novel decoding algorithms, GraphAr delivers a comprehensive and efficient solution for optimizing LPG-specific operations.
} 
%This tailored approach significantly improves label filtering performance, whether it involves simple or complex conditions.

Our key contributions can be summarized as follows:
\vspace{-0.5mm}
\begin{itemize}
   \setlength{\itemsep}{-1.5pt}
   \item Elucidation of challenges and limitations in existing tabular formats for managing LPGs in data lakes (Section~\ref{sec:background}).
   \item A strategic choice of Parquet compatibility, a standardized YAML to fully express LPG semantics, and detailed specification for organizing LPGs in Parquet (Section~\ref{sec:overview}).
   \item Development of specialized optimization techniques for enhancing performance in neighbor retrieval and label filtering operations (Sections~\ref{sec:format} and \ref{sec:label}). 
   %, both  are innovative additions, to our knowledge, not previously applied within the existing file formats or graph systems.
   These are built upon Parquet's advanced encoding features and are complemented by two innovative and efficient decoding algorithms.
   \item Comprehensive performance evaluation of \gar~compared to Parquet and Acero-based implementations, highlighting substantial speed gains: on average $4452\times$ for neighbor retrieval, $14.8\times$ for label filtering, and $29.5\times$ for end-to-end workloads. 
   And the potential for integrating \gar~into existing graph systems (Section~\ref{sec:evaluation}).
\end{itemize}
\section{Background and Key Challenges}
\label{sec:background}
In this section, we discuss the limitations of using tabular file formats like Parquet and ORC in data lakes for LPGs, a common graph data model. We explore how these formats inadequately support LPG representation and efficient graph queries, laying the groundwork for the challenges that \gar tackles. %, outlined at the section's end.

\subsection{Tabular Formats in Data Lakes}
\label{sec:tabular}
Tabular data is key to data lakes, aiding efficient organization, analysis, and data extraction from large sets. Columnar formats like Parquet~\cite{parquet} and ORC~\cite{orc} are popular due to their robust features. 
Unlike row-based formats such as CSV, %they offer key advantages. 
they allow faster queries by enabling selective column reading, avoiding unnecessary data.
Additionally, they offer diverse and efficient compression and encoding strategies, such as delta encoding to compress the variance between consecutive values, and run-length encoding to compress repetitive values.
These techniques not only reduce storage needs but also enhance processing speeds.
Another advantage of Parquet and ORC is predicate pushdown, which enhances query performance by moving filters closer to the storage layer, thus minimizing reads.

The combination of selective column reading, efficient compression, and predicate pushdown positions Parquet and ORC as the go-to choices for managing tabular data in data lakes.
Previous studies have demonstrated the importance of leveraging their capabilities for optimizing relational data management~\cite{floratou2014sql, ivanov2020impact, DeltaLake}.
Recent research~\cite{li2023selection, Vectorizing2013, simd4, FlexPushdownDB, PushdownDB, AWS_S3_Select} has also explored enhancements to tabular formats, utilizing CPU instruction sets like BMI and SIMD.
In this paper, we will focus on Parquet, but the techniques discussed can be seamlessly adapted to other columnar formats such as ORC.

\begin{figure}[tbp]
  \centering
  \includegraphics[width=\linewidth]{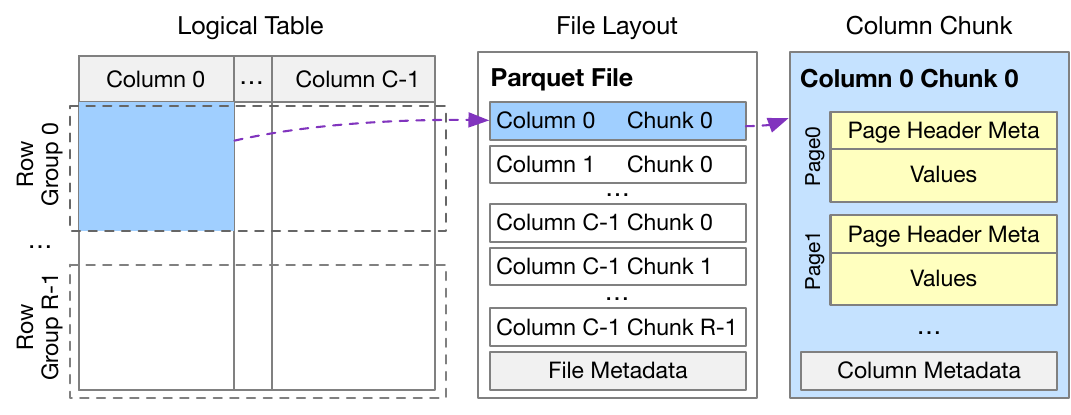}
  \vspace{-7mm}
  \caption{The internal structure of a Parquet file for a logical table with $C$ columns and $R$ row groups.}
  \vspace{-2mm}
  \label{fig:parquet}
\end{figure}

\stitle{Parquet.} %Parquet files store data in columns, each accessible independently, as depicted in Figure~\ref{fig:parquet}.
Figure~\ref{fig:parquet} illustrates the internal structure of a Parquet file. 
Structurally, a Parquet file represents a table, organized into row groups for logical segmentation. % without enforcing a physical layout. 
Within a row group, the data of a column is stored in a column chunk, which is guaranteed to be contiguous in the file.
Column chunks are further divided into pages, the indivisible units for compression and encoding.
These pages, which can vary in type, are interleaved in a column chunk. 
%Structurally, a Parquet file represents a table organized into one or more row groups, with each group containing a single column chunk per column. Each column chunk then comprises one or more pages.

Parquet files contain three layers of metadata: file metadata, column metadata, and page header metadata.
The file metadata directs to the starting points of each column's metadata.
Inside the column chunks and pages, the respective column and page header metadata are stored, offering a detailed description of the data.
This includes data types, encoding, and compression schemes, facilitating efficient and selective access to data pages within columns.

\subsection{Labeled Property Graphs}
\label{sec:lpg}
Labeled Property Graphs (LPGs)~\cite{angles2023pg, baken2020linked, anikin2019labeled} excel at representing complex relationships and semantics in a natural manner. Their flexible schema allows for accommodating the diverse and evolving nature of big data within repositories, making them integral to data lakes.
LPG serves as the canonical data model in many graph systems~\cite{neo4j, gonzalez2014graphx, TigerGraph} and graph query languages~\cite{pgql, cypher, cypher_paper, gremlin, GQL, pattern}, enabling queries and analytics to uncover valuable insights and patterns.
Formally, an LPG is defined as \( G = (V, E, T_V, T_E, P, L) \), where \( V \) is a set of vertices, \( E \) a set of edges, \( T_V \) and \( T_E \) the types of vertices and edges respectively, \( P \) the properties, and \( L \) the labels.

\begin{figure}[tbp]
  \centering
  \begin{subfigure}{\linewidth}
    \centering
    \includegraphics[width=\linewidth]{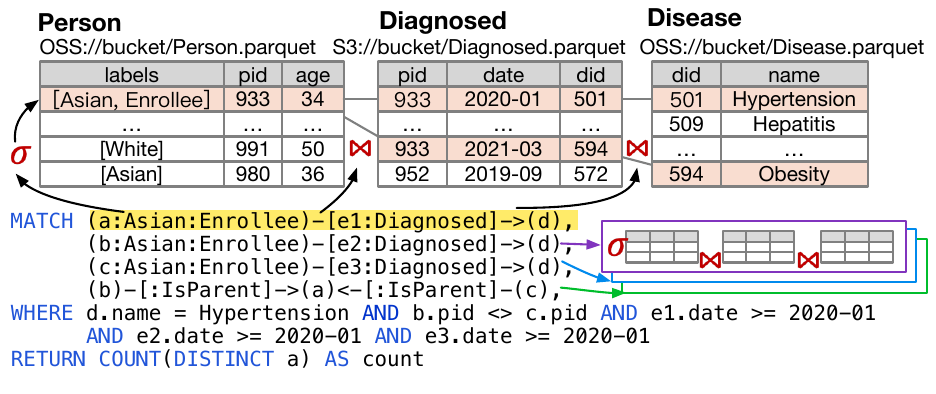}
    \vspace{-7mm}
    \caption{Execution workflow of the query on tabular formats.}
    \label{fig:intro}
  \end{subfigure}

  \hfill
  \vspace{1mm}

  \begin{subfigure}{\linewidth}
    \centering
    \includegraphics[width=\linewidth]{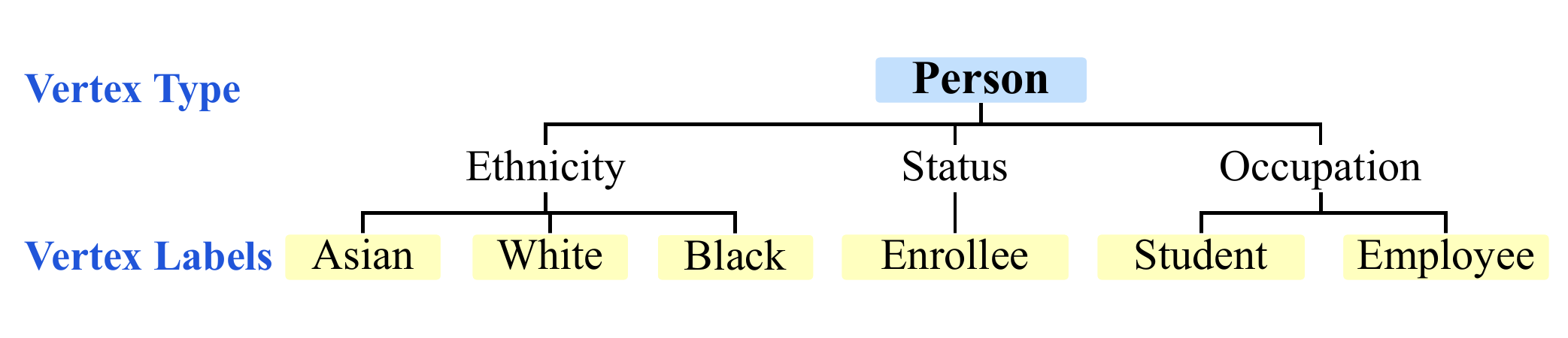}
    \vspace{-7mm}
    \caption{\revise{Inherent multi-labeling characteristics inside Person vertices.}}
    \label{fig:vertex-labels}
  \end{subfigure}
  \caption{An example of querying LPGs on tabular formats.}
  \vspace{-2mm}
  \label{fig:background}
\end{figure}

For each vertex \( v \in V \) and edge \( e \in E \), they are associated with a type \( t_v \in T_V \) and \( t_e \in T_E \) respectively, and can have optional properties. A property \( p \in P \) is specific to a vertex or an edge type, with a unique identifier within its type and a pre-defined datatype for its values. This implies that vertices or edges of the same type share the same set of properties, and can be stored in a tabular format, as shown in Figure~\ref{fig:intro}.

Furthermore, each label \( l \in L \) has a unique identifier, usually a string. Each vertex type \( t_v \) is linked with a set of candidate labels \( L(t_v) \subseteq L \), allowing each vertex of this type to be assigned zero or more labels. 
\revisenote{M5 R3.D4}
\revise{
Labels hold significant importance in LPGs as they represent classifications and characteristics of entities, while properties serve as attributes to store additional information.
LPGs allow vertices to have multiple labels, offering a flexible and expressive way to describe entities. For instance, Figure~\ref{fig:vertex-labels} illustrates an inheritance hierarchy of labels within the \emph{Person} vertices, where a vertex can be labeled as both Asian and Enrollee.}
Although edge types could technically also be labeled, we focus solely on vertex labels in this paper, aligning with common graph query practice\footnote{Graph query languages like Cypher and Gremlin typically adhere to the convention that an edge can have only one classification, corresponding to the edge type in LPG model. Nevertheless, the strategies for vertex labels discussed in this paper can be seamlessly extended to support edge labels.}.

%\footnote{Both Gremlin and Cypher follow the convention that an edge can have one and only one (string) label, which corresponds to the edge type in the LPG model. Edge labels are relative rarity in real-worlds. Edge label filtering is not supported by most graph query languages and it can be more efficiently handled as a post-neighbor retrieval filtering condition. Nevertheless, our strategy can be seamlessly extended to support edge label filtering.}.

While Parquet is highly effective for storing individual vertex and edge types \( T_V \) and \( T_E \) along with their associated properties due to the columnar structure and data compression capabilities, it falls short in capturing the interconnected schema essential for linking vertices with edges, e.g., to express the relationships across the three tables of Figure~\ref{fig:intro}, which represent two vertex types and one edge type.
This limitation is crucial for efficient graph traversal and pattern matching. Moreover, Parquet lacks native support for the multi-labeling capability of LPGs, resulting in a loss of complex semantics and relationships inherent to LPGs.

%\subsection{Limitations in Supporting Graph Queries}
\subsection{Querying Labeled Property Graphs}
The core feature shared among graph query languages is the facility for pattern matching~\cite{pattern, cypher}. %, essential for exploring complex relationships between entities. 
%This feature enables users to effectively inquire and investigate the complex connections and interdependencies within a graph.
This capability allows for in-depth analysis of the relationships between entities, uncovering valuable insights and patterns that may not be readily apparent in other data models.
Given the fact that a LPG consists of topology, labels, and properties, a graph pattern is then defined as vertices and their connections through edges, filtered based on labels and property values~\cite{GQL}.
Figure~\ref{fig:intro} illustrates the example workload mentioned in Figure~\ref{fig:example}, expressed in Cypher.
And the steps for matching \lstinline[language=SQL]{(a:Asian:Enrollee)-[e1:Diagnosed]->(d)} are highlighted in the figure.
%Figure~\ref{fig:intro} illustrates a simplified GQL query on the example LPG, to identify persons labeled as Asian and Enrollee who are between the ages of 30 and 50 and have been diagnosed with some disease.
%The query then returns the top $10$ diseases, ordered in descending order based on the average age of the patients.
%Figure~\ref{fig:intro} illustrates a simplified GQL example involving two vertex types and one edge type. 
%A simplified GQL example is illustrated in Figure~\ref{fig:intro}.

When it comes to properties, a viable approach is to use native tabular data, leveraging existing formats for efficient storage and property-related operations. 
%This leverages the storage efficiency of existing formats, while optimizing property-related operations in graph queries. 
However, this approach struggles with two crucial aspects of pattern matching.

\sstab
Firstly, tabular formats lack native support for representing graph topology, making it difficult to efficiently fetch the neighboring vertices and edges for a given vertex. A common workaround is to store edge endpoints as properties and use the join operations across multiple tables to retrieve neighbors, as shown in Figure~\ref{fig:intro}. 
However, this approach is often inefficient due to the computational overhead of multiple joins.

\sstab
%Secondly, LPGs allow vertices to have multiple labels, offering a flexible and expressive way to describe entities. 
Secondly, label filtering is a unique feature in graph queries, to enable the selection of specific subsets of vertices, making it an essential element in all graph query languages~\cite{GQL, cypher, gremlin, pgql}.
Existing formats do not natively accommodate this flexibility and do not provide a foundation for label-based optimizations. Encoding labels as ordinary properties and performing filtering by string matching, as seen in the initial step of Figure~\ref{fig:intro}, limits the expressive power of vertex representation and hampers efficient label handling. 
%, which is a feature unique to graph queries.

These query-side inefficiencies highlight the limitations of using tabular formats for graphs. %While they are proficient in storing and querying tabular data, they are inadequate for the specialized requirements of graph querying. 
The reliance on suboptimal workarounds, such as multiple joins for topology expansion and makeshift encoding schemes for label filtering, compromises query performance and complicates the query process. This paves the way for the challenges we seek to address. % in the following sections.

\subsection{Key Challenges Addressed by \gar}
\label{sec:challenges}
The development of \gar is motivated by the specific limitations of existing tabular formats for both representing LPGs and supporting efficient graph queries. %Below, we outline the key challenges that \gar aims to tackle, each corresponding to a dedicated section later in this paper.

\stitle{Challenge 1: Effective LPG representation.}
LPGs use type-based organization and specific label/property definitions to form a cohesive graph structure. This enables precise and targeted querying. Existing tabular formats fall short of capturing these intricate semantics, necessitating a specialized solution. This challenge is addressed in Section~\ref{sec:overview}.

\stitle{Challenge 2: Efficient neighbor retrieval.}
A fundamental aspect of graph queries is the operation known as \emph{neighbor retrieval}. This is vital for quick access to adjacent vertices and edges, thus accelerating graph traversal. Existing tabular formats, however, do not natively or efficiently support this crucial operation. This issue is tackled in Section~\ref{sec:format}.

\stitle{Challenge 3: Optimized label filtering.}
\emph{Label filtering} is a primary filtering mechanism in graph queries, allowing for the early elimination of irrelevant data. Existing tabular formats do not natively support this operation, making it a ripe area for optimization. This is the focus of Section~\ref{sec:label}.

Each of these challenges represents a gap in the capabilities of current tabular formats for graph data.
They serve as the focus areas for the technical contributions of \gar, with each corresponding to a dedicated section in this paper.
% detailed in the subsequent sections.

\begin{figure*}[tbp]
  \centering
  \begin{subfigure}{0.23\linewidth}
    \centering
    \includegraphics[width=\linewidth]{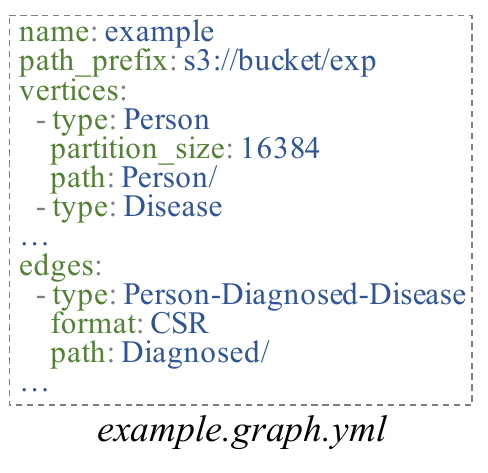}
    \vspace{-7mm}
    \caption{Metadata.}
    \label{fig:overview-a}
  \end{subfigure}
  \hfill
  \begin{subfigure}{0.42\linewidth}
    \centering
    \includegraphics[width=\linewidth]{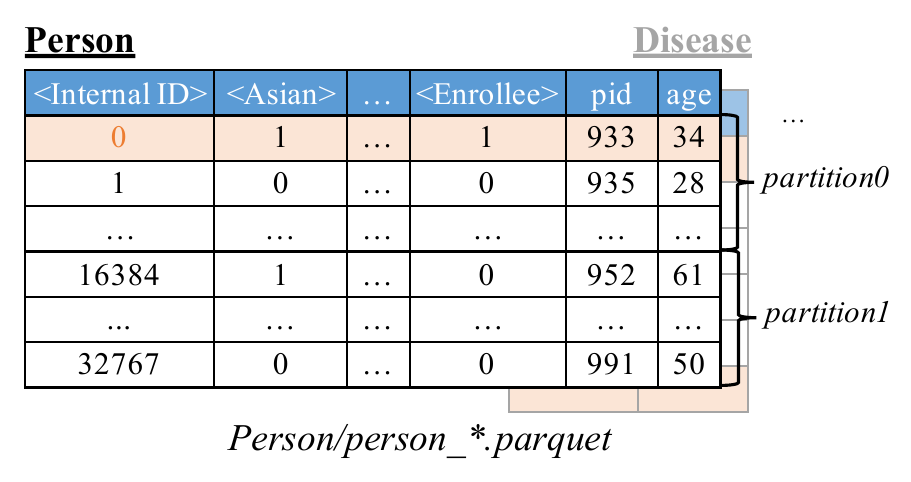}
    \vspace{-7mm}
    \caption{Vertex table of Person (and Disease).}
    \label{fig:overview-b}
  \end{subfigure}
  \hfill
  \begin{subfigure}{0.34\linewidth}
    \centering
    \includegraphics[width=\linewidth]{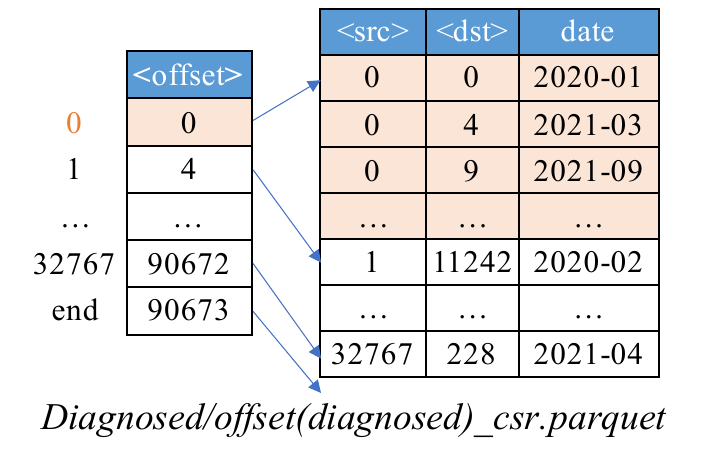}
    \vspace{-7mm}
    \caption{Edge table of Person-Diagnosed-Disease.}
    \label{fig:overview-c}
  \end{subfigure}
  \caption{The metadata and data layout for the example graph in \gar.}
  \vspace{-1mm}
  \label{fig:overview}
\end{figure*}

%\section{Representing LPG Semantics in Data Lakes}
\section{Representing LPGs in \gar}
\label{sec:overview}
This section provides an overview of how \gar customizes the representation of LPGs in data lakes.
It begins by outlining its goals and non-goals, providing clarity on the rationale behind its design.
Next, it explains the strategies employed for data organization and layout. %, emphasizing the significance of schema metadata and the use of Parquet as the payload format.
Lastly, the section describes how \gar seamlessly integrates into the data lake ecosystem.
%Initially, it highlights the significance of schema metadata in \gar, essential for capturing the multifaceted, type-based schema of LPGs. It then explicates the strategies employed for data organization and layout, emphasizing the use of Parquet as the payload file format. Lastly, the section describes how \gar seamlessly integrates into the data lake ecosystem.

%\subsection{Schema Metadata for LPGs}
%\label{sec:schemametadata}

\subsection{Goals and Non-Goals}
\label{sec:goals}
\stitle{Goals.} \gar's primary goal is to provide an efficient storage and management scheme for LPGs in data lakes, specifically targeting the three main challenges outlined in Section~\ref{sec:background}. %, to enhance both storage and query capabilities.

\sstab
\gar also seeks compatibility with both data lake and graph processing ecosystems for smooth integration with a variety of existing tools and systems.

\stitle{Non-Goals.} \gar does not intend to replace existing data lake formats like Parquet and ORC, but to maximize their benefits and offer additional features for LPGs.

\sstab
Also, \gar is not a graph computing engine; rather, it can be non-intrusively integrated with  graph processing systems, either serving as the archival format or acting as a data source.

\sstab
In line with the established practices of data warehousing and lake house architectures, both Parquet/ORC and \gar adhere to data immutability norms, treating batch-generated data as immutable once created.
%Data is generated in batches and treated as immutable once created. 
Higher-level systems such as graph databases manage mutation (e.g., adding, deleting, or updating vertices) through specialized, non-standardized file and in-memory versioning methods.

\subsection{Data Organization and Layout}
In \gar, vertices and edges are organized according to their types, which aligns with the principles of the LPG model.
Parquet is utilized as the payload file format for storing the data in the physical storage, while YAML files are used to capture schema metadata.

\stitle{Schema metadata.}  A YAML file (Figure~\ref{fig:overview-a}) stores the metadata for a graph. It specifies important attributes such as file path prefixes and vertex/edge types. This file serves as a nimble yet effective way to capture metadata that is not accommodated by Parquet, while Parquet files include specific details about properties and labels within their internal schemas. It can optionally include partition sizes, allowing for data to be segmented into multiple physical Parquet files, thereby enabling parallelism at the file level.

\stitle{Vertex table.} As depicted in Figure~\ref{fig:overview-b}, each row in the vertex table represents a unique vertex, identified by a 0-indexed internal ID, stored in the \emph{<Internal ID>} column. 
%Optionally, 
When partitioning is enabled, for the $i$-th partition, its internal IDs start at $\text{partition\_size} \times i$, and within each partition, IDs are sorted in ascending order.
Bubbles\footnote{``Bubbles'' refer to the allowance for some ranges of internal IDs or edge segments not to correspond to any vertices or edges.}  are allowed at the end of each partition, meaning the actual number of rows can be less than or equal to the partition size.

Property columns (\emph{pid} and \emph{age}) are named after their respective properties and hold the corresponding values with specified datatypes.
In terms of labels, a set of candidate labels is defined for each vertex type.
Then a vertex can have an arbitrary number of labels from the corresponding set.
For example, the vertex type \emph{Person} may have labels to represent ethnicity.
%may have three candidate labels: \emph{pm} (for premium member), \emph{spm} (for super premium member), and \emph{sus} (for suspicious).
For efficient storage and filtering of labels, \gar~uses a binary representation to maintain each label in an individual column named with angle brackets, e.g., \emph{<Asian>} and \emph{<Enrollee>}.
Additionally, advanced encoding/decoding techniques are applied, which will be discussed in Section~\ref{sec:label}.

\stitle{Edge table.} Edges are also organized and stored in Parquet files, similar to vertices.
Figure~\ref{fig:overview-c} showcases the layout of the edge table for type \emph{Person-Diagnosed-Disease}, where \emph{Person} and \emph{Disease} represent the source and destination vertex types, while \emph{Diagnosed} signifies the classification of the relationships. Each edge is associated with the internal IDs of its source and destination vertices, stored in columns named \emph{<src>} and \emph{<dst>}.
Edge properties, and optional partitioning, are handled in the same way as vertices.
%Properties are handled in the same way as vertices, and partitioning is also allowed for edges.
%Properties are stored in columns named according to their property names. 
%Partitioning is also allowed in edges in a similar way to vertices.
%The source (and destination) vertices are referred to as internal IDs and maintained in the \emph{<src>} (and \emph{<dst>}) columns, while edge properties are handled in the same way as vertices.

\stitle{Optimized access patterns for neighbor retrieval.}
The layout strategy in \gar leverages the columnar storage capabilities of Parquet to facilitate efficient graph traversal. Edges are sorted first by source vertex IDs and then by destination vertex IDs. This sorting strategy optimizes various access patterns.
For row-wise access, the layout closely resembles the Coordinate List (COO) format, making it well-suited for edge-centric operations. On the other hand, an auxiliary index table, denoted as \emph{<offset>}, is introduced to enable more efficient vertex-centric operations.
The \emph{<offset>} table aligns with the partitions in the vertex table, and when applied to the source vertices, facilitates retrieval patterns similar to Compressed Sparse Row (CSR).
%This separate index table accommodates scenarios where a vertex has multiple types of edges created separately from different data sources, a common occurrence in data lakes.
Likewise, a similar approach can be applied to enable Compressed Sparse Column (CSC)-like access. \gar allows for efficient bidirectional neighbor retrieval through two sorted tables for the same edge type.
CSR, CSC, and COO are widely adopted for representing graphs, thus \gar ensures compatibility with existing graph-related systems.

These layout strategies are complemented by encoding and decoding optimizations (Section~\ref{sec:format}). Collectively, these strategies enhance both the data management and query capabilities of \gar.

\subsection{Incorporation with Data Lakes}
The design of \gar makes it especially well-suited for integration with data lakes, largely due to its reliance on widely adopted standards such as Parquet and YAML.
% Parquet for data storage and YAML for graph metadata management.

\stitle{Data transformation and construction.}
The \gar format is essentially a specialized layout of Parquet files accompanied by a YAML metadata file. This enables the use of existing data processing frameworks like Apache Spark, Acero, and Hadoop, which can access various graph systems like Neo4j, TigerGraph and Nebula, or other types of database systems through their respective connectors. These frameworks can also ingest a multitude of data formats including logs, relational tables, JSON, and more. Such flexibility provides users with the ability to construct, transform, and store LPGs in data lakes from a wide array of data sources. %To simplify the process of generating files in the \gar format, we also provide a Spark library specifically designed for this purpose.

\stitle{Downstream system integration.}
Since \gar is fundamentally based on Parquet and YAML, it is straightforward to use it as a data source for downstream systems. Many systems already have the capability to ingest Parquet files, making \gar a convenient and efficient data storage scheme.

\stitle{Graph-specific optimizations.}
In addition to serving as a flexible storage format, \gar is also optimized for graph-specific operations. These optimizations, include advanced query pushdown techniques and other performance enhancements that are particularly useful for graph-specific tasks and queries within data lakes (see more from Sections~\ref{sec:format} and~\ref{sec:label}).
%\section{Optimizing Neighbor Retrieval}
\section{Efficient Neighbor Retrieval}
\label{sec:format}
In this section, we address the critical challenge of efficient neighbor retrieval in graphs. We leverage Parquet's data pages and introduce page-aligned collections (PAC) for streamlined neighbor identification.
Additionally, we utilize Parquet's delta encoding to enhance performance, and introduce an innovative decoding strategy that leverages BMI and SIMD. 
%Additionally, we utilize Parquet's delta encoding as an efficiency-boosting technique.
%We then introduce a novel decoding strategy that leverages BMI and SIMD. 
%, using bitmaps as the data structure for collections in PAC.
All of these techniques are integrated into \gar, resulting in highly efficient neighbor retrieval.
%A fundamenb stal and frequently encountered operation in graph queries involves the accessing of neighboring vertices associated with a given target vertex. 
%The accessing operation entails retrieving the internal IDs of the neighboring vertices and subsequently selecting the corresponding properties and labels from the vertex table according to these IDs. 
%A pioneering solution proposed in~\cite{li2023selection} underscores that a bitmap representation of IDs enables selection pushdown in columnar storages. 

%In practice, as the vertex table is organized as miniblocks in parquet, we use a miniblock-aligned bitmap defined as follows:

%\subsection{Workflow and Challenges}
\subsection{Workflow of Neighbor Retrieval}

%In data lake environments, storage systems like Parquet often use data pages to match the data storage with the access granularity of underlying storage layers, such as block devices or cloud storage services like Amazon S3.
Parquet use data pages to match the data storage with the access granularity of the underlying storage layers, with a page represents the minimum unit of data that can be read from or written to the storage layer, as illustrated in Figure~\ref{fig:parquet}.
And encoding and decoding are applied at the page level.

For LPG queries, a common operation is to retrieve specific property values of neighboring vertices, given a queried vertex, e.g., obtaining the \emph{name} values of \emph{Disease} vertices connected to a particular \emph{Person} vertex.
Assuming the CSR format utilized for storing edge table \emph{Person-Diagnosed-Disease}, the typical workflow for this operation involves: 1) Using the \emph{<offset>} index and \emph{<dst>} column of the edge table to identify and fetch the first relevant page from the target vertex table (a page from the \emph{name} column in the \emph{Disease} table). This page contains at least one neighboring vertex pertinent to the query, and may also include other irrelevant vertices;
2) Selectively fetching the property values corresponding to the neighboring vertices within that page. This step is repeated iteratively for each subsequent page containing the targeted neighbors.

\begin{figure}[tbp]
  \centering
  \begin{subfigure}{.85\linewidth}
    \centering
    \includegraphics[width=\linewidth]{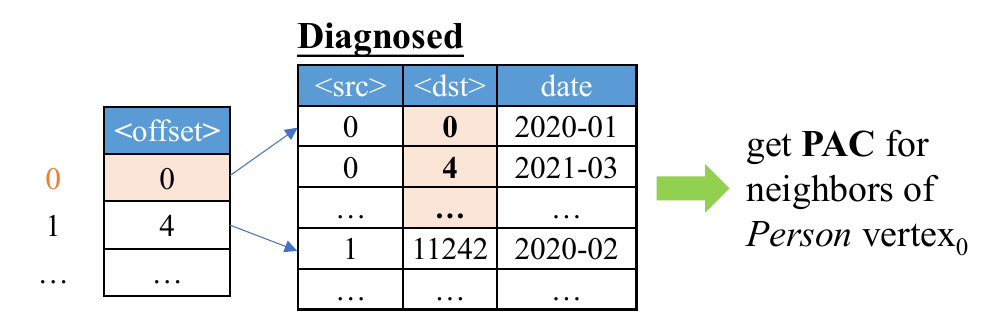}
    \vspace{-7mm}
    \caption{\revise{Neighbor retrieval of a source vertex to obtain PAC.}}
    \label{fig:pac-a}
  \end{subfigure}

  \hfill
  \vspace{1mm}

  \begin{subfigure}{.85\linewidth}
    \centering
    \includegraphics[width=\linewidth]{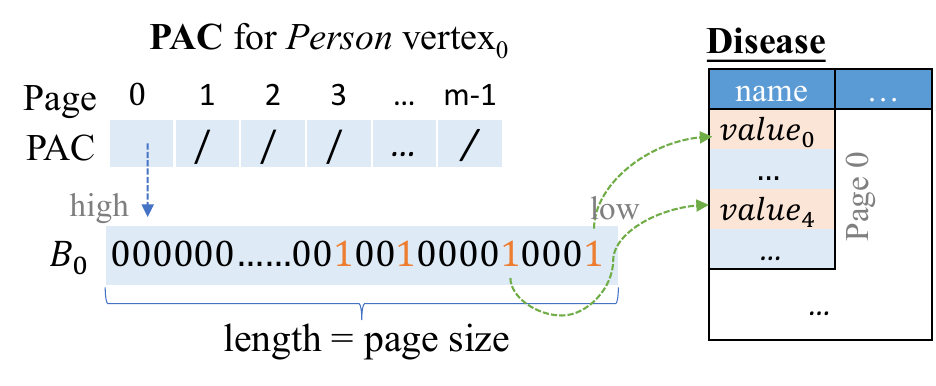}
    \vspace{-8mm}
    \caption{\revise{Using the PAC to get the properties of its neighbors.}}
    \label{fig:pac-b}
  \end{subfigure}
  \caption{\revise{An example of PAC and its usage.}}
  \vspace{-3mm}
  \label{fig:pac}
\end{figure}

This workflow highlights the two primary steps during neighbor retrieval. The first is to identify which pages in the vertex table contain the neighboring vertices relevant to the query. The second is to fetch the relevant property values within each of these pages efficiently.  Firstly, we formalize the neighbor retrieval operation:

\revise{
  \revisenote{M5 R3.W2 R3.D2}
  \begin{definition}[Neighbor Retrieval]
    \label{def:neighbor-retrieval}
    Given a vertex \( v \), the operation of neighbor retrieval returns a data structure \( \mathcal{C} \) representing the internal IDs of the neighboring vertices connected to \( v \).
  \end{definition}
  \vspace{-0.8ex}

  \stitle{Requirement of $\mathcal{C}$.} The data structure \( \mathcal{C} \) should facilitate the retrieval of the relevant property values of neighboring vertices, while minimizing space and processing overhead.
  %Bitmaps are a natural choice for the first requirement but their space demands could be a concern for the second requirement.
  Accordingly, we introduce the concept of page-aligned collections (PAC).

  \subsection{Page-Aligned Collections}
  \label{sec:pac}

  \begin{definition}[Page-aligned collections (PAC)]
    \label{def:collections}
    Given a column in vertex table that includes \( m \) pages, the PAC \( \mathcal{C} = [C_0, \ldots, C_{m-1}] \) is a list of up to \( m \) collections. Each \( C_i \) stores a set of internal IDs in the corresponding page. Non-empty collections in \( \mathcal{C} \) are retained, while empty ones are omitted.
  \end{definition}

  \vspace{-0.6ex}  
  \stitle{Intuition.} Each collection \( C \) in PAC returned by neighbor retrieval corresponds to a data page of the target vertex table.
  To save space and avoid unnecessary processing, empty collections are omitted.
  This is based on the sparsity and locality of real-world graphs, which often results in irrelevant pages.
  %For the second task, 
  Subsequently, the internal IDs within each collection enable the retrieval of only the relevant property values through a \emph{selection} process.
  The remaining challenges then involve 1) optimizing each collection's representation for quick value retrieval and 2) efficiently generating the PAC $\mathcal{C}$.

A pioneering solution~\cite{li2023selection} underscores the transformation of indices into a bitmap representation to enable selection pushdown in columnar storages.
To adopt this approach to efficiently retrieve the properties of neighbors, which addresses the first challenge, we adopt a bitmap representation $B$ for each \emph{non-empty} collection $C$ in PAC, where $B[i]=1$ indicates the existence of $i$-th element.

\revisenote{M5 R3.W2 R3.D3}
Figure~\ref{fig:pac-a} illustrates the neighbor retrieval of a source vertex to obtain a PAC, and Figure~\ref{fig:pac-b} demonstrates the usage of PAC to get the properties of its neighbors.
For this illustrated example, only $C_0$ is non-empty, and $B_0$ is the bitmap representation of $C_0$.
%Figure~\ref{fig:pac} illustrates a PAC and its bitmap representation, where only $C_1$ is non-empty, and $B_1$ is the bitmap representation of $C_1$.
The bitmap representation can be used to facilitate the selection pushdown of vertex properties or labels, e.g., fetching the properties of \emph{name} in the target vertex table \emph{Disease}, for the neighbors of \emph{Person} $\text{vertex}_0$.
For the second challenge of efficiently generating PAC (i.e., Figure~\ref{fig:pac-a}), we utilize Parquet's delta encoding and a novel decoding strategy, which we detail in the following.
}

\eat{
  \begin{definition}[Miniblock-aligned Bitmap]
    \label{def:bitmap}
    Given a vertex table include $m$ miniblock, the miniblock-aligned bitmap $B=[\mathcal{B}_0,\dots,\mathcal{B}_{m-1}]$ include $m$ miniblock bitmap. As many miniblock bitmaps may be empty, i.e., all bits are 0, the non-empty miniblock bitmap in $B$ is retained, and the empty miniblock is omitted.
  \end{definition}
  %Formally, we define the neighbor retrieval operation as follows:
  \begin{definition}[Neighbor Retrieval]
    \label{def:neighbor-retrieval}
    Given a vertex $v$, the neighbor retrieval operation returns the Miniblock-aligned Bitmap $B$ representing the internal IDs of the neighboring vertices connected to $v$.
  \end{definition}
  To facilitate our demonstration, we consider a logical bitmap $B$ and denote $B[u]=1$ indicating $u$ is the neighbor of $v$.

  In Section~\ref{sec:overview}, we introduce the data layout for graph topology in \gar, which is based on a CSR-like (or CSC-like) format~\footnote{In this section, we focus on the CSR format, while the CSC format can be implemented in a similar way.}.
  In this section, we first explore the potential of combining data layout and delta encoding to reduce the data load overhead.
  Additionally, we propose an efficient decoding methodology that leverages the potential of BMI and SIMD, further improving the performance of neighbor retrieval.
}

\subsection{Delta Encoding}
To compute the PAC $\mathcal{C}$, the encoded internal IDs of neighboring vertices need to be loaded, sourced from the edge table.
In a data lake scenario, where data can be stored remotely, the loading process can be more time-consuming than processing due to I/O limitations.
%Consequently, the loading of internal IDs from the edge table is a more time-consuming process than processing. 
To address this issue, we investigate the use of delta encoding for data compression, consequently reducing data load volume.

% In our pursuit of optimizing the storage and retrieval efficiency of graph topology data, we explore the encoding technique used in \gar: the integration of delta encoding. 
\stitle{Delta encoding.} %Previous research, including Gemini~\cite{zhu2016gemini} and Facebook-Graph~\cite{ugander2011anatomy}, has demonstrated that real-world graphs often exhibit both sparsity and locality. 
Previous research~\cite{zhu2016gemini, ugander2011anatomy} has demonstrated that real-world graphs often exhibit both sparsity and locality.
This means that while a vertex's neighbors might be spread across the entire graph, they are more likely to cluster within certain ID ranges.
Such patterns arise from various factors, such as the inherent clustering in real-life graphs, where vertices within a cluster are more interconnected, and the methods used for data collection (e.g., crawlers or the organic/viral growth patterns of social networks like Facebook or TikTok).
Systems~\cite{boldi2004webgraph,zhu2016gemini} have utilized such sparsity and locality to enable efficient partitioning or compression.

In \gar, such inherent locality, reinforced by our meticulous dual-key sorting of edges in the carefully designed layout, to enable incremental arrangement of internal IDs for a vertex's neighbors, serves as the basis for delta encoding, which is highly effective for both the \emph{<src>} and the \emph{<dst>} columns in the edge table.
%This approach takes advantage of the carefully designed layout we have established, where edges are sorted meticulously. This strategic sorting enables the incremental arrangement of internal vertex IDs for the neighbors of a given vertex, which serves as the basis for our delta encoding technique.
Delta encoding works by storing the deltas between consecutive values rather than each value separately. The deltas, which often have small values, can be stored more compactly, requiring fewer bits.
% In our context, this technique provides significant benefits by greatly reducing the storage demands of vertex IDs in the edge table, for both source and destination columns.
% It also contributes to the efficiency of data retrieval, as the data to be loaded is reduced. % Additionally, the decoding process can be further accelerated through the use of BMI and SIMD, as detailed in the next section.

\eat{
  \begin{figure}[tbp]
    \centering
    \includegraphics[width=0.95\linewidth]{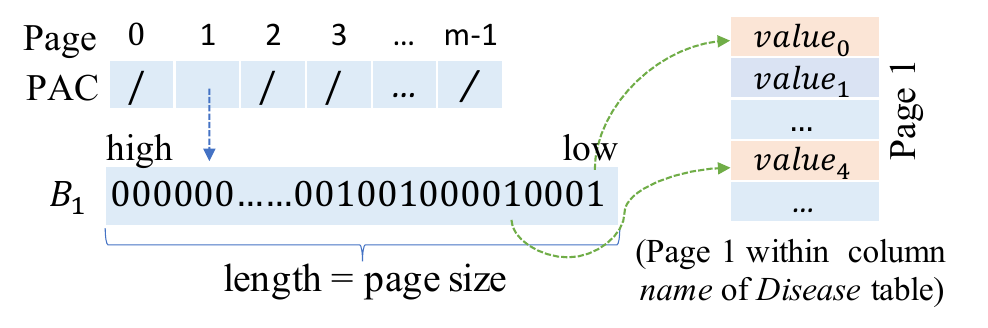}
    \vspace{-4mm}
    \caption{An example of PAC and the bitmap representation.}
    %\vspace{-1mm}
    \label{fig:pac}
  \end{figure}

  \begin{figure*}[tbp]
    \centering
    \includegraphics[width=.95\linewidth]{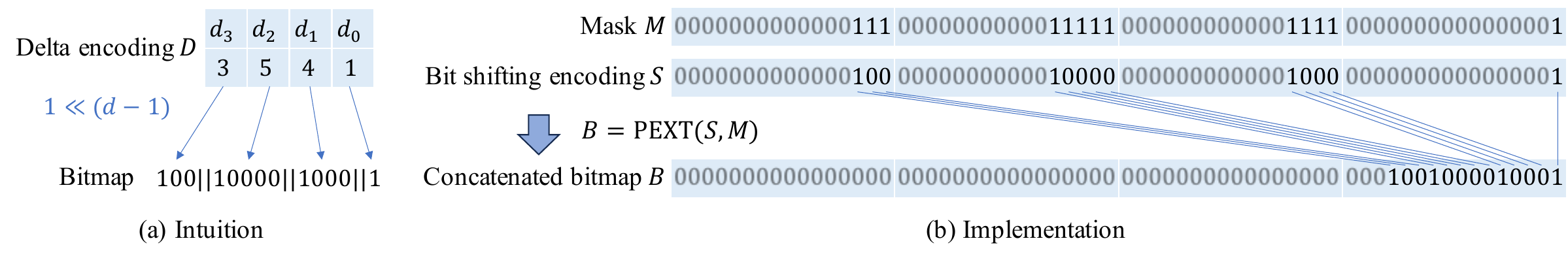}
    %\vspace{-2mm}
    \caption{An example of accelerating decoding via BMI.}
    %\vspace{-5mm}
    \label{fig:BMI}
  \end{figure*}
}

\begin{figure*}[tbp]
  \centering
  \begin{subfigure}{0.25\linewidth}
    \centering
    \includegraphics[width=\linewidth]{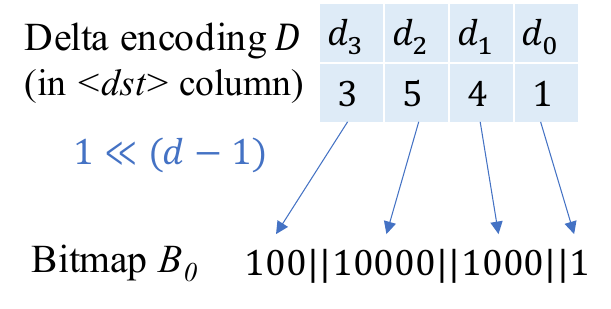}
    \vspace{-7mm}
    \caption{Intuition.}
    \label{fig:BMI-intuition}
  \end{subfigure}
  \hfill
  \begin{subfigure}{0.7\linewidth}
    \centering
    \includegraphics[width=\linewidth]{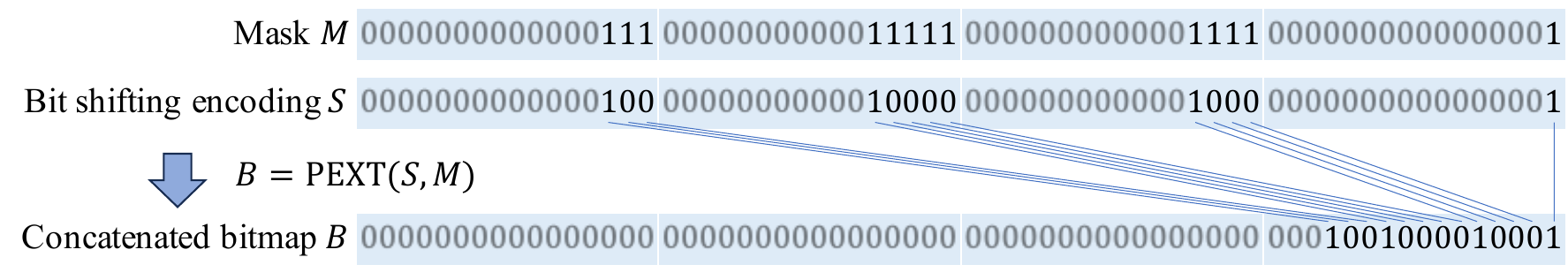}
    \vspace{-7mm}
    \caption{Implementation.}
    \label{fig:BMI-implementation}
  \end{subfigure}
  \vspace{-1mm}
  \caption{\revise{An example of accelerating decoding via BMI.}}
  \vspace{-2mm}
  \label{fig:BMI}
\end{figure*}

\stitle{Implementations.} %Both \parquet~and ORC, utilized as payload file formats in \gar, come with inherent support for delta encoding.
%In this paper, we take \parquet~as an example to implement and evaluate our approach.
%Our implementation is based on Parquet, which offers built-in support for delta encoding.
We utilize Parquet's built-in support for delta encoding~\cite{Lemire_2013}, which is implemented based on miniblocks.
%, adapted from the binary packing approach~\cite{Lemire_2013}.
%This technique, adapted from the binary packing approach described in~\cite{Lemire_2013}, comprises a header and blocks of delta encoded values. 
%Each block is composed of miniblocks, with each miniblock being binary packed using its own bit width.
%We adhere to the default miniblock size of 32 values and the default block size of 128 values (i.e., 4 miniblocks).
Each miniblock (with a size of $32$ values) is binary packed using its own bit width, which should be a power of 2 for data alignment purposes.
%and for data alignment purposes, the bit width should be a power of $2$.
%For data alignment purposes, the bit width of a miniblock must be a power of 2.
This design allows us to adapt to changes in the data distribution, as the bit width of each miniblock is dynamically adjusted to minimize storage consumption.
According to our evaluation across various real-world graphs, as detailed in Section~\ref{sec:evaluation-neighbor-retrieval},
%the delta encoding technique can reduce the expected loaded data volume (which can be measured by storage consumption) by $58.1\%$ to $81.0\%$ compared to without delta encoding.
%As a result, it brings an individual speedup of up to $3.6\times$ for neighbor retrieval. %, compared to without delta encoding.
the delta encoding technique can reduce the expected loaded data volume by $58.1\%$ to $81.0\%$ compared to without delta encoding.
As a result, it brings an individual speedup of $2.7\times$ for neighbor retrieval.

\eat{
  \stitle{Quantitative analysis.} Figure~\ref{fig:opt1-motivation} provides a quantitative analysis to demonstrate the effectiveness of delta encoding in \gar.
  We use four representative real-world graphs of varying sizes, with statistics provided in Table~\ref{tab:datasets}, and re-organize the edges according to the CSR-like format.
  The analysis focuses on the vertex with the largest outgoing degree in each graph, which corresponds to the vertex with the largest number of neighbors in the edge table.
  To maintain the internal IDs of the neighbors for the vertex, using the delta encoding described above, the number of bits required for each miniblock is calculated.
  The percentage of miniblocks with a bit width not exceeding 1 bit, 2 bits, and so on, is then determined and depicted in Figure~\ref{fig:opt1-motivation}.
  The results clearly demonstrate the effectiveness of delta encoding in reducing storage consumption, as the majority of miniblocks have a significantly lower bit width than 32 bits.
  Experiments in Section~\ref{sec:evaluation} further evaluate the overall storage efficiency.
  On average, \gar~only requires $27.3\%$ of the space for comprehensive graph topology data maintenance compared to without delta encoding.

  \begin{figure}[tbp]
    \centering
    \includegraphics[width=0.9\linewidth]{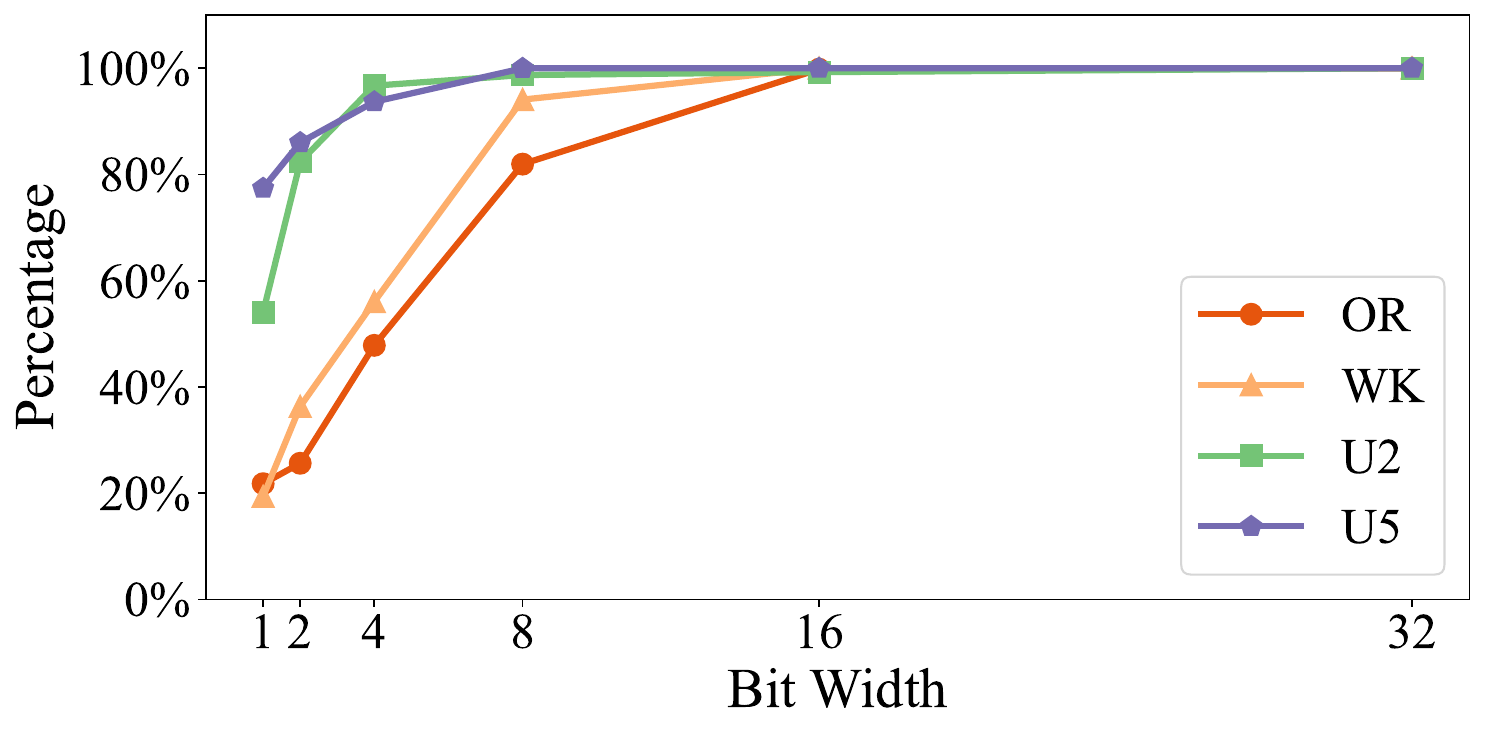}
    %\vspace{-2mm}
    \caption{Distribution of miniblock bit widths.}
    %\vspace{-2mm}
    \label{fig:opt1-motivation}
  \end{figure}
}

\subsection{BMI-based Decoding}
\label{sec:bmi-decoding}
\stitle{Challenges.} Whiles delta encoding effectively reduces loading costs, it introduces additional decoding computation.
% Furthermore, it is noteworthy that the destination column within the edge table exclusively holds the internal ID of the destination vertex, serving as an intermediary step in the process of indexing within the vertex table.
% This facilitates the retrieval of essential properties or labels associated with neighboring vertices.
% A pioneering solution proposed in~\cite{li2023selection} underscores the transformation of indices into a bitmap representation to enable selection pushdown in columnar storages.
% The authors advocate for a technique grounded in BMI (Bit Manipulation Instructions), which further expedites the decoding process through the utilization of the bitmap. 
% To adapt this technique to our context, to support the selection pushdown of vertex properties and labels, 
% we engineer an efficient decoding process from delta-encoded format to bitmap representation, in which BMI plays a pivotal role.
%Fortunately, as indicated in \cite{li2023selection}, the BMI (Bit Manipulation Instructions) can be used to accelerate the bit-level operation.
Some existing works~\cite{li2023selection, Lemire_2013, simd4, simd1, simd2, simd3} have explored the use of BMI (Bit Manipulation Instructions) and SIMD (Single Instruction, Multiple Data) to accelerate the data compression, decoding, scanning, or management.
% We can leverage the capabilities of SIMD in conjunction with BMI to further improve the performance. 
However, delta encoding involves data dependencies that make vectorization challenging.
The decoding of the $(i+1)$-th neighbor depends on the prior decoding of the $i$-th neighbor.

\eat{
Another unsolved issue is the choice of implementing collection $C$ in PAC.
A pioneering solution proposed in~\cite{li2023selection} underscores the transformation of indices into a bitmap representation to enable selection pushdown in columnar storages.
To adopt this approach to efficiently retrieve the properties of neighbors, which constitutes the second task of neighbor retrieval, we adopt a bitmap representation $B$ for each \emph{non-empty} collection $C$ in PAC, %, while omitting empty collections.
%Specifically, we denote $B$ as a bitmap representation of a collection $C$, 
where $B[i]=1$ indicates the existence of $i$-th element.
Figure~\ref{fig:pac} illustrates a PAC and its bitmap representation, where only $C_1$ is non-empty, and $B_1$ is the bitmap representation of $C_1$.
The bitmap representation can be used to facilitate the selection pushdown of vertex properties or labels, e.g., fetching the properties of \emph{name} in the target vertex table \emph{Disease}, for the neighbors of a \emph{person} vertex.
}

\revisenote{M5 R3.W2 R3.D3}
\revise{In our context, the critical challenge is to generate the bitmap representation of PAC efficiently from the delta-encoded neighbor IDs, which are stored in the \emph{<dst>} column of the edge table (as shown in Figure~\ref{fig:BMI-intuition}).}
Existing techniques are not suitable for our context due to the data dependencies involved.
However, by taking advantage of the sophisticated instruction sets offered by modern CPUs, we can exploit the functionalities of BMI together with SIMD operations to overcome this challenge, through an innovative decoding strategy.
%Next, let us delve into the intuitive understanding of how to efficiently generate a bitmap from delta encoding.

\stitle{Intuition.} To ensure clarity, we initially consider a simplified scenario where each delta value is compressed to a 4-bit size.
Conventionally, a two-step approach is used to decode the delta-encoded data, in which the current encoded value is added to the previously decoded ID to obtain the current ID, and then the bitmap is updated bit by bit based on the decoded IDs.
However, our analysis reveals that this two-step process is redundant.
By leveraging the bit-shifting encoding $1<<(d-1)$ for each delta value $d$,
we can generate the bitmap by concatenating the bit-shifting encodings: $1<<(d_{n-1}-1)||\dots||1<<(d_1-1)||1<<(d_0-1)$,  where $d_0,d_1,\dots,d_{n-1}$ represent $n$ delta values, and $||$ represents the concatenation operator.
This principle is visually depicted in Figure~\ref{fig:BMI-intuition}.

\stitle{Acceleration via BMI and SIMD.} In practical implementation, the bit-shifting encoding is maintained using a fixed-length datatype, characterized by zero-padding on the left side.
In our example, 16 bits are sufficient to accommodate the 4-bit deltas.
The bit-shifting encodings of 4 values are stored in a 64-bit register, allowing for parallel generation and processing using SIMD instructions.
Then, the focus shifts to the compaction of these encodings. Fortunately, the Parallel Bit Extract (PEXT) operation, a specialized CPU instruction in BMI, facilitates efficient aggregation of discrete bits from source positions into contiguous bits within the destination, governed by a selector mask.
This process is illustrated in Figure~\ref{fig:BMI-implementation}.
%The process of this operation is illustrated in Figure~\ref{fig:BMI-implementation}.

The subsequent challenge is to generate the required mask, achieved by deriving the $i$-th mask $m_i$ from the $i$-th bit-shifting encoding $s_i$ using the equation $m_i=(s_i<<1)-1$.
\revisenote{R1.W2}\revise{The advantage of this operation is its potential for convenient parallel execution, facilitated by direct manipulation of the mask sequence residing within a 64-bit register.}
Specifically, $M$ can be acquired via the following steps: 1) a bitwise (in our example, 16 bits) shift of each $s_i$ to the left by 1 bit, parallelized through SIMD instructions like
\lstinline[basicstyle=\small]{_mm_slli_epi16};
and 2) a bitwise subtraction of 1 from the result of the previous step, which can be accelerated via instructions like \lstinline[basicstyle=\small]{_mm_sub_epi16}.
All these SIMD instructions utilized in our implementations are widely available in modern CPUs, included in SSE2 (which we use) and more recent sets such as AVX2 and AVX-512.

In general, vectorization demonstrates greater efficiency when the bit width is smaller, as it allows for more significant parallelism.
Our extensive evaluations have confirmed that the BMI-based decoding approach outperforms the default decoding approach in Parquet when the bit width is within 4 bits, with performance improvements ranging from $3.3\%$ to $110\%$.
Therefore, we utilize this BMI-based approach for miniblocks with a bit width of 1 to 4 bits, while resorting to the default delta decoding of \parquet~for larger bit widths.
The combination of data layout, delta encoding, and this adaptive decoding strategy results in an advanced topology management paradigm, enabling efficient neighbor retrieval.

%\section{Optimizing Label Filtering}
\section{Optimized Label Filtering}
\label{sec:label}
Labels serve as a representation of the classification or characteristics of vertices in a graph.
Filtering vertices by labels is a fundamental syntax in graph query languages, as it allows querying specific subsets of vertices.
%\footnote{Edge labels are relative rarity in real-worlds. Edge label filtering is not supported by most graph query languages and it can be more efficiently handled as a post-neighbor retrieval filtering condition. Nevertheless, our strategy can be seamlessly extended to support edge label filtering.}, 
Existing approaches~\cite{neo4j_csv, neo4j_spark} of fitting graphs into tabular data often treat labels as regular properties, encapsulating them within a string or list, as seen in Figure~\ref{fig:intro}.
This approach overlooks the inherent differences between labels and properties, leading to inefficient label filtering, due to the need for decoding string representations and conducting string matching.

%In this section, we delve into the distinction between labels and properties.
%Recognizing the widespread use of labels as filter conditions and their unique nature, we begin our exploration by analyzing a simplified condition.
%We then develop a specialized format for labels leveraging binary representation and run-length encoding (RLE).
Recognizing the widespread use of labels as filter conditions and their unique nature, we develop a specialized format for labels leveraging binary representation and run-length encoding (RLE), for handling simple conditions.
To support complex conditions introduced by user-defined functions that involves multiple labels, we enhance our methodology with a novel merge-based decoding algorithm, further improving efficiency and adaptability.

\subsection{Handling Simple Conditions}
We start by considering the simple condition that focuses on the existence of a single label. In essence, the existence or absence of a label can be effectively represented using binary notation, where the value 1 indicates the existence of the label and 0 indicates its absence, as demonstrated in Figure~\ref{fig:overview-b}. This binary representation offers two significant advantages: 1) it reduces the computational burden associated with decoding and matching as well as simplifies the filtering process as follows; 2) it enables efficient compression.

%\vspace{-1ex}
\begin{definition}[Simple Condition Filtering]
  \label{def:simple label existence condition}
  Given a label $l$ and an existence/absence indicator $e$, the simple condition label filtering returns the PAC $\mathcal{C}$, where
  %\vspace{-1ex}
  \begin{equation}
    \left\{
    \begin{aligned}
      v \in \mathcal{C}, \text{ if } v.label[l]=e \\
      v \notin \mathcal{C}, \text{ if } v.label[l]\neq e
    \end{aligned}
    \right.
    %B[v]=\left\{
    %  \begin{aligned}
    %  & 1,  \text{ if } v.label[l]=e \\
    %  & 0,  \text{ otherwise}
    %  \end{aligned}
    %  \right.
  \end{equation}
\end{definition}
%\vspace{-1ex}
% \stitle{Binary representation.} 

\stitle{Encoding.} To compress consecutive runs of 0s or 1s, we utilize the technique of run-length encoding (RLE), which represents them as a single number.
This run-length format naturally transforms the binary representation of a label into an interval-based structure.
We then adopt a list $P$ to define the positions of intervals.
The $i$-th interval is represented by $[P[i], P[i+1])$, where $P[i]$ refers to the $i$-th element within $P$.
Besides, it is required to record whether the vertices of the first interval $[P[0], P[1])$ contain the label or not, i.e., the first value.
By leveraging this technique, the storage consumption of labels can be significantly reduced. %, as existing labels in real-world graphs are often sparse.

\stitle{Decoding.} Beyond efficient compression, the RLE approach seamlessly accommodates the decoding requirements for filter conditions.
Specifically, to filter vertices with (or without) a specific label, we can simply select all odd intervals or all even intervals from the list $P$, based on the condition and the first value, instead of evaluating each vertex individually.
It reduces the time complexity from $O(n)$ to $O(|P|)$, where $n$ represents the number of vertices and $|P|$ represents the size of the interval list $P$.
In real-world graphs, $|P| \ll n$ is often observed, due to the sparsity of labels and natural clustering of vertices with similar labels.

\subsection{Extending to Complex Conditions}
Expanding beyond the realm of simple label existence, we encounter the intricacies of dealing with complex conditions involving multiple labels.
Consider a scenario where we need to find vertices with specific label combinations, such as the GQL pattern \lstinline[language=SQL]{MATCH (person:Asian&Enrollee)} (or in Cypher, \lstinline[language=SQL]{MATCH (person:Asian:Enrollee)}), which retrieves vertices labeled as Asian and Enrollee.
A more complex pattern can be \lstinline[language=SQL]{MATCH (person:(Asian&!Enrollee)|Student)}, which retrieves vertices labeled as Asian but not Enrollee, or labeled as Student.
To handle such scenarios, we employ user-defined functions (UDFs) to represent complex filter conditions.
The UDF $f$ takes a vertex $v$ as input and returns a boolean value $f(v)$, indicating whether the vertex satisfies the condition or not.
Formally, we define the complex condition filtering as follows.

%\vspace{-1.5ex}
\begin{definition}[Complex Condition Filtering]
  \label{def:complex label existence condition}
  Given a UDF $f$, the filtering returns the PAC $\mathcal{C}$, where
  %\vspace{-1.5ex}
  \begin{equation}
    \left\{
    \begin{aligned}
      v \in \mathcal{C}, \text{ if }  f(v)=\text{true} \\
      v \notin \mathcal{C},  \text{ if }  f(v)=\text{false}
    \end{aligned}
    \right.
    %      B[v]=\left\{
    %        \begin{aligned}
    %        & 1,  \text{ if } f(v)=\text{true} \\
    %        & 0,  \text{ otherwise}
    %        \end{aligned}
    %        \right.
  \end{equation}
\end{definition}
%\vspace{-1.5ex}

The intuitive approach would be to tackle each vertex independently, decoding RLE into the binary representation. % detailed earlier.
However, directly evaluating the UDF for every vertex proves impractical, as it retains the same complexity as the most straightforward approach.

\stitle{Intuition.} Inspired by the concept of discretization, two key questions arise: 1) Can we solely evaluate the condition for one representative vertex within each interval? %, thereby reducing the time complexity from $O(n)$ to $O(|P|)$, where $|P|$ signifies the size of the interval list $P$? 
2) How can we efficiently identify these intervals where the encompassed vertices share the same labels?
The affirmative answer to the first question emerges through the following theorem:
%\vspace{-2ex}
\begin{theorem}
  Consider $k$ interval lists $P_0,P_1,\dots,P_{k-1}$, where the vertices in $[P_i[j],P_i[j+1])$ share the same value for the $i$-th label. If an iterval $[s,e)$ is not broken by any position, i.e.,
  %If there is no position that breaks the interval $[s,e)$, i.e.,
  %\vspace{-1.5ex}
  \begin{equation}
    \nexists i\in[0,k),j\in[0,|P_i|) \quad s < P_i[j] < e,
  \end{equation}
  the vertices within the interval $[s,e)$ have the same labels, i.e.,
  %\vspace{-1.5ex}
  \begin{equation}
    \forall u,v\in[s,e),l\in[0,k) \quad u.label[l]=v.label[l].
  \end{equation}
\end{theorem}
%As the interval $[s,e)$ remains unbroken by the existing intervals in list $P$, indicating that any two vertices within the interval share the same value, no matter for which label, 
Consequently, it is sufficient to call the UDF for the vertex $s$ alone, as for any vertex $v$ in the interval $[s,e)$, $v$ and $s$ share the same labels, thus $f(v)=f(s)$.

\begin{figure}[tbp]
  \centering
  \includegraphics[width=\linewidth]{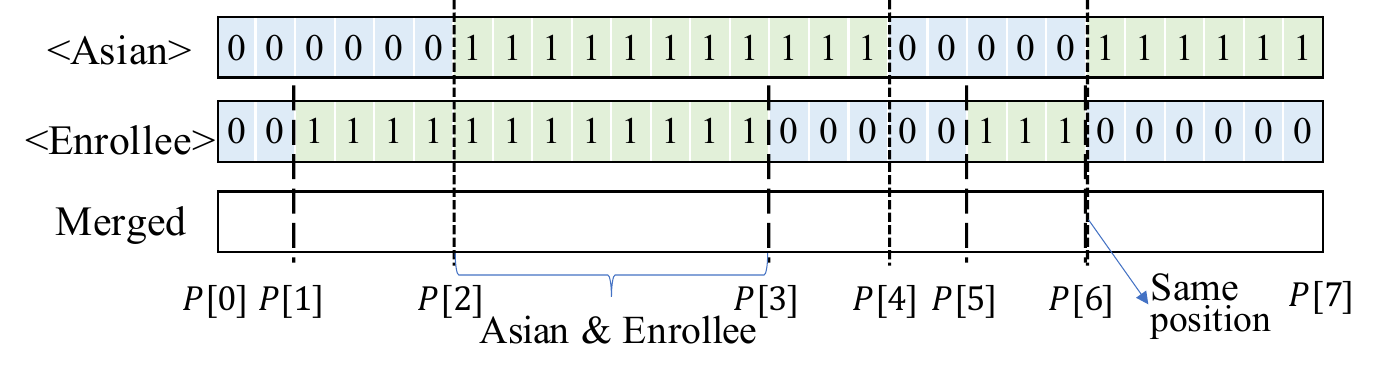}
  \vspace{-8.5mm}
  \caption{\revise{An example of merging intervals.}}
  \vspace{-2.5mm}
  \label{fig:merge}
\end{figure}

\stitle{Merge-based algorithm.} Partitioning an interval into multiple segments proves unnecessary and counterproductive as it would escalate complexity.
Therefore, our focus narrows down to intervals formed by existing positions, which also addresses the second question.
To obtain the exact intervals, we can sort the positions in all interval lists $P_0, P_1,\dots, P_{k-1}$.
This sorting can be accelerated by leveraging the inherent order within the $k$ lists, allowing for seamless merging of $k$ sorted lists into one list $P$.
\revisenote{M5 R3.W2 R3.D3}
\revise{Figure~\ref{fig:merge} demonstrates an example of interval determination for a complex condition containing two labels, Asian and Enrollee.}
Within the interval $[P[i],P[i+1])$, the vertices share identical labels, necessitating the UDF to be invoked solely for one representative vertex.
Additionally, the presence of position $P[6]$ for both labels underscores the importance of merging to avoid redundant computations.

By employing innovative label treatment, interval-based encoding/decoding, and complex condition handling, \gar~is able to achieve highly efficient label filtering.
%and filter pushdown, improving the management of LPGs.
%These advancements solidify \gar~as a robust solution for storing and managing LPGs.

\section{Evaluation}
\label{sec:evaluation}
In this section, we evaluate \gar on a range of graphs, through micro-benchmarks and end-to-end graph query workloads. 
%In this section, we evaluate \gar on a range of real-world and synthetic graphs, through micro-benchmarks focusing on neighbor retrieval and label filtering, as well as end-to-end graph query workloads. 
We also illustrate the potential benefits \gar offers to current graph processing systems. %highlighting its capacity to elevate performance or expand applicability. 
Our findings validate \gar as an efficient storage scheme for LPG storage and querying in data lakes.

\begin{figure*}[tbp]
    \centering
    \begin{subfigure}{0.405\linewidth}
        \centering
        \includegraphics[width=\linewidth]{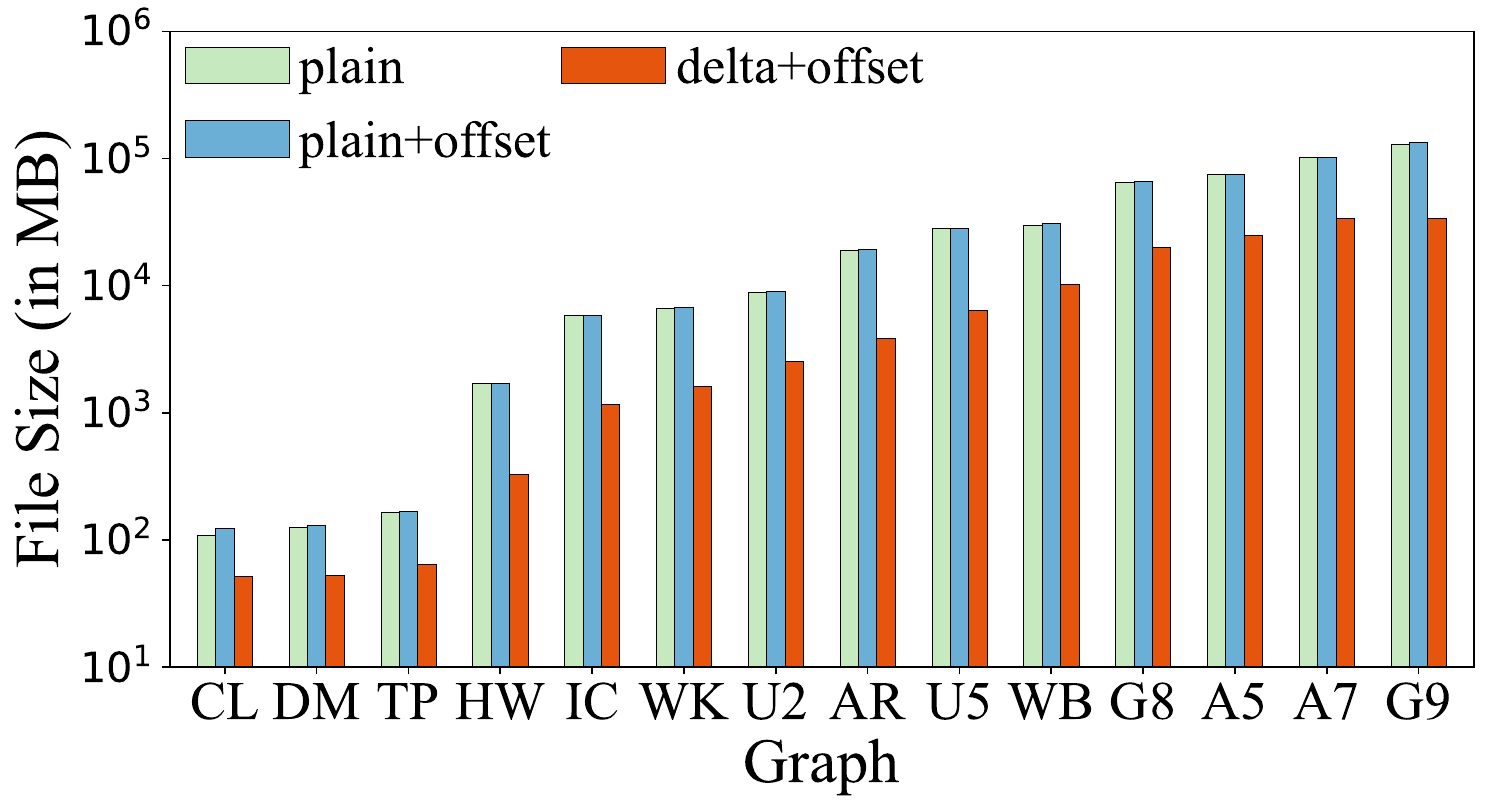}
        \vspace{-7mm}
        \caption{\revise{Storage consumption for topology.}}
        \label{fig:opt1-encoding}
    \end{subfigure}
    \hfill
    \begin{subfigure}{0.405\linewidth}
        \centering
        \includegraphics[width=\linewidth]{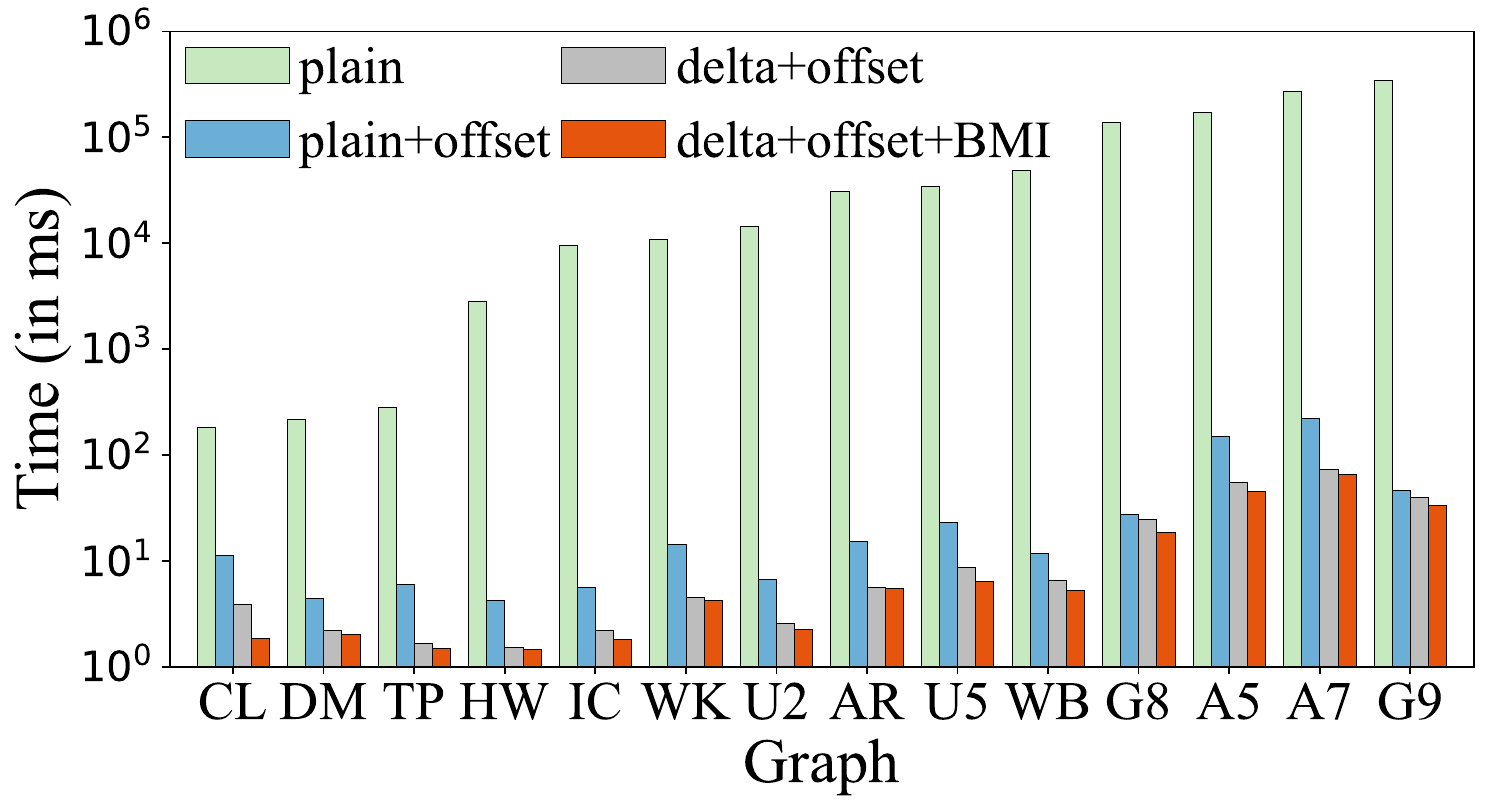}
        \vspace{-7mm}
        \caption{\revise{Neighbor retrieval time.}}
        \label{fig:opt1-decoding}
    \end{subfigure}
    \hfill
    \begin{subfigure}{0.18\linewidth}
        \centering
        \includegraphics[width=\linewidth]{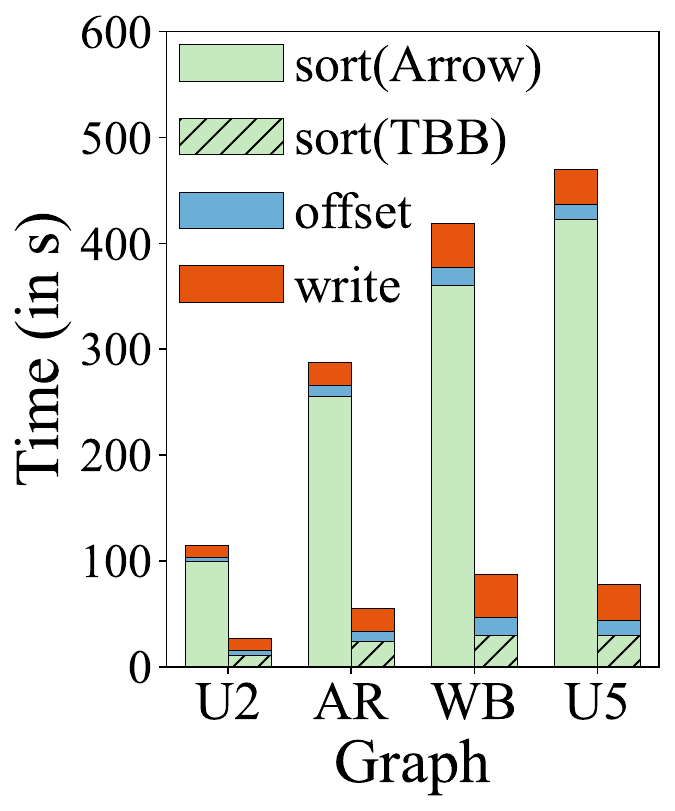}
        \vspace{-7mm}
        \caption{\revise{Transformation time.}}
        \label{fig:opt1-generation}
    \end{subfigure}
    \caption{Micro-benchmark of neighbor retrieval.}
    \vspace{-1mm}
\end{figure*}

\begin{table}[t]
    \centering
    \caption{\revise{Statistics of the graphs in our evaluation.}}
    \vspace{-2mm}
    \label{tab:datasets}
    \fontsize{8}{9}\selectfont
    \begin{tabular}{|l|l|r|r|}
        \hline
        \textbf{Abbr.} & \multicolumn{1}{c|}{\textbf{Graph}} & \multicolumn{1}{c|}{$\mathbf{|V|}$} & \multicolumn{1}{c|}{$\mathbf{|E|}$} \\
        \hline
        \revise{A5} & \revise{Alibaba synthetic (scale 5)} & \revise{75.0M} & \revise{4.93B} \\
        \revise{A7} & \revise{Alibaba synthetic (scale 7)} & \revise{100M} & \revise{6.69B} \\
        AR & arabic-2005~\cite{matrix} & 22.7M & 1.27B \\%& 576K \\
        BL & bloom~\cite{neo4j_dataset} & 33.0K & 29.7K \\
        CF & com-friendster~\cite{snapnets} & 65.6M & 1.81B \\
        CI & citations~\cite{neo4j_dataset} & 264K & 221K \\
        CL & cont1-l~\cite{nr} & 1.92M & 7.03M \\%& 1.28M \\
        DM & degme~\cite{nr} & 659K & 8.13M \\%& 624K \\
        \revise{G8} & \revise{Graph500-28~\cite{graph500}} & \revise{268M} & \revise{4.29B}\\
        \revise{G9} & \revise{Graph500-29~\cite{graph500}} & \revise{537M} & \revise{8.59B}\\
        HW & hollywood-2009~\cite{matrix} & 1.14M & 113M \\% & 11.5K \\
        OL & icij-offshoreleaks~\cite{neo4j_dataset} & 1.97M & 3.27M \\
        PP & icij-paradise-papers~\cite{neo4j_dataset} & 163K & 364K \\
        IC & indochina-2004~\cite{matrix} & 7.41M & 384M \\%& 256K \\
        %LR & LargeRegFile~\cite{nr} & 2.11M & 4.94M \\%& 656K \\
        %LG & legis-graph~\cite{neo4j_dataset} & 11.8K & 523K \\
        NM & network-management~\cite{neo4j_dataset} & 83.8K & 181K \\
        AX & ogbn-arxiv~\cite{hu2020ogb} & 169K & 1.17M \\
        MA & ogbn-mag~\cite{hu2020ogb} & 736K & 21.1M \\ 
        OS & openstreetmap~\cite{neo4j_dataset} & 71.6K & 76.0K \\
        %OR & orkut~\cite{snapnets} & 3.07M & 213M \\%& 33.3K \\
        PO & pole~\cite{neo4j_dataset} & 61.5K & 105.8K \\
        SF30 & SNB Interactive SF-30~\cite{LDBC} & 99.4M & 655M \\
        SF100 & SNB Interactive SF-100~\cite{LDBC} & 318M & 2.15B \\
        SF300 & SNB Interactive SF-300~\cite{LDBC} & 908M & 6.29B \\
        TP & tp-6~\cite{nr} & 1.01M & 10.7M \\%& 445K \\
        %TV & twitter-v2~\cite{neo4j_dataset} & 43.3K & 56.4K \\
        TT & twitter-trolls~\cite{neo4j_dataset} & 281K & 493K \\
        U2 & uk-2002~\cite{matrix} & 18.5M & 589M \\%& 195K \\
        U5 & uk-2005~\cite{matrix} & 39.5M & 1.85B \\%& 1.78M \\
        WB & webbase-2001~\cite{matrix}& 118M & 2.01B \\%& 816K \\
        WK & wiki~\cite{KONECT} & 13.6M & 437M \\%& 1.05M \\
        \hline
    \end{tabular}
    \vspace{-4mm}
\end{table}

\subsection{Experimental Setup}

\stitle{Platform.} If not otherwise mentioned, our experiments are conducted on an Alibaba Cloud r6.6xlarge instance, equipped with a 24-core Intel(R) Xeon(R) Platinum 8269CY CPU at 2.50GHz and 192GB RAM, running 64-bit Ubuntu 20.04 LTS. The data is hosted on a 200GB PL0 ESSD with a peak I/O throughput of 180MB/s. 
Additional tests on other platforms and S3-like storage yield similar results. 
For timing metrics, we use single-threaded executions and report either average or distribution times based on multiple runs for accuracy.
Exceptionally, the integration experiments utilize a cluster of 8 separate instances to emulate a distributed environment.

\stitle{Baselines.} \gar~is developed in C++ on Apache Arrow~\cite{arrow}, an open-source, high-performance library that supports columnar formats like Parquet and ORC.
For the micro-benchmarks, we compare \gar~against Arrow/Parquet (version 13.0.0), due to the popularity and high-performance of Parquet.
Both \gar and the baseline follow Parquet's default configurations, which include a row group length of \(1024 \times 1024\) and a 1MB page size. 
%While there are other related works in this area, we discuss them in Section~\ref{sec:related} and do not directly compare them with \gar, since they are either not applicable or demonstrate worse performance than Parquet in the context of data lakes.
\revisenote{M4 R2.W3 R2.D2}
\revise{
For end-to-end workloads, we compare \gar~against widely-used frameworks including Apache Acero~\cite{Acero}, Apache Pinot~\cite{pinot} and Neo4j (Community 5.21.0)~\cite{neo4j}. 
%Apache TinkerPop~\cite{tinkerpop}. 
%Other related works are discussed in Section~\ref{sec:related}.
}
In the integration experiments, we incorporate \gar into GraphScope~\cite{graphscope}, a widely-used graph processing system, and compare its integrated performance against GraphScope's original implementation.

\revisenote{M1 R1.W1}
\revise{
\stitle{Datasets.} As summarized in Table~\ref{tab:datasets}, our evaluation includes a variety of graphs, span different sizes and domains, including social networks and web graphs. We also use synthetic graphs generated by data generators of the LDBC SNB~\cite{LDBC} and Graph500~\cite{graph500}, both of which are widely recognized benchmarks.
Additionally, we utilize graphs (A5 and A7) generated that closely mimic the characteristics of graphs in the e-commerce production environment at Alibaba.
%, designed to mimic real-world graph characteristics. 
%They are part of the LDBC Social Network Benchmark~\cite{LDBC,ldbc-bi}, and we use its query set for end-to-end workload assessments.
}

\subsection{Micro-Benchmark of Neighbor Retrieval}
\label{sec:evaluation-neighbor-retrieval}
We evaluate \gar's optimizations in neighbor retrieval through micro-benchmarks on selected graphs characterized by a large edge set (\( |E| \)). Our results substantiate its efficacy in enhancing storage efficiency and retrieval performance.

%\subsubsection{Storage Efficiency}
\stitle{Storage efficiency.}
We compare \gar~with baseline methods by measuring the storage consumed by encoded \textit{Parquet} files that store the graph's topological data. Two baseline approaches are considered: 1) ``plain'', which employs plain encoding for the source and destination columns, and 2) ``plain + offset'', which extends the ``plain'' method by sorting edges and adding an offset column to mark each vertex's starting edge position. 
As Figure~\ref{fig:opt1-encoding} depicts, the inclusion of offsets results in a modest increase in storage requirements, with space usage growing by $0.5\%$ to $14.8\%$, as the number of vertices is typically much smaller than the number of edges.

\gar~leverages delta encoding for source and destination columns and plain encoding for offsets. %, adhering to default Arrow/\textit{Parquet} settings. 
\revisenote{M1 R1.W1}
\revise{
The result is a notable storage advantage: on average, it requires only 29.2\% of the storage needed by the baseline ``plain + offset''.
%The result is a notable storage advantage: on average, it requires only 27.3\% of the storage needed by the baseline ``plain + offset''.
} This efficiency in storage is particularly beneficial for query performance, as data lake queries are often I/O-bound. The transition from storage efficiency to retrieval performance is elaborated further in the next experiment.

%\subsubsection{Performance of Neighbor Retrieval}
\stitle{Performance of neighbor retrieval.}
To evaluate \gar's efficiency in neighbor retrieval, we query vertices with the largest degree in selected graphs, maintaining edges in CSR-like or CSC-like formats depending on the degree type. 
\revisenote{M1 R1.W1 R3.D1}
\revise{
%Figure~\ref{fig:opt1-decoding} shows that \gar significantly outperforms the baselines, achieving speedups ranging from \(49.8\times\) to \(9232.4\times\) over the ``plain'' method and \(2.1\times\) to \(6.1\times\) over the ``plain + offset'' baseline. 
%These gains are attributed to the offset integration and delta encoding, as well as our BMI-based decoding.
%The offset integration alone accounts for a speedup of between $11.0\times$ and $4177.2\times$, and delta encoding contributes an additional $1.8\times$ to $3.6\times$ speedup on top of that.
%While not explicitly shown in the log-scaled Figure~\ref{fig:opt1-decoding}, the use of BMI and SIMD further enhances performance, with an average improvement of \(20.1\%\).
Figure~\ref{fig:opt1-decoding} shows that \gar significantly outperforms the baselines, achieving an average speedup of \(4452\times\) over the ``plain'' method, \(3.05\times\) over ``plain + offset'', and \(1.23\times\) over ``delta + offset''.
% and \(3.1\times\) over the ``plain + offset'' baseline.
These gains are attributed to the offset integration and delta encoding, as well as our BMI-based decoding.
The offset integration alone accounts for an average speedup of $1993\times$, and delta encoding provides an additional $2.48\times$ speedup.
Our innovative decoding method, which leverages BMI and SIMD, further enhances performance within this optimized context, achieving a $1.23\times$ speedup on top of ``delta + offset''.
}
%While not explicitly shown in the log-scaled Figure~\ref{fig:opt1-decoding}, the use of BMI and SIMD further enhances performance, with an additional improvement of \(23.2\%\) on average.

\begin{figure*}[tbp]
    \centering
    \begin{subfigure}{0.48\linewidth}
        \centering
        \includegraphics[width=\linewidth]{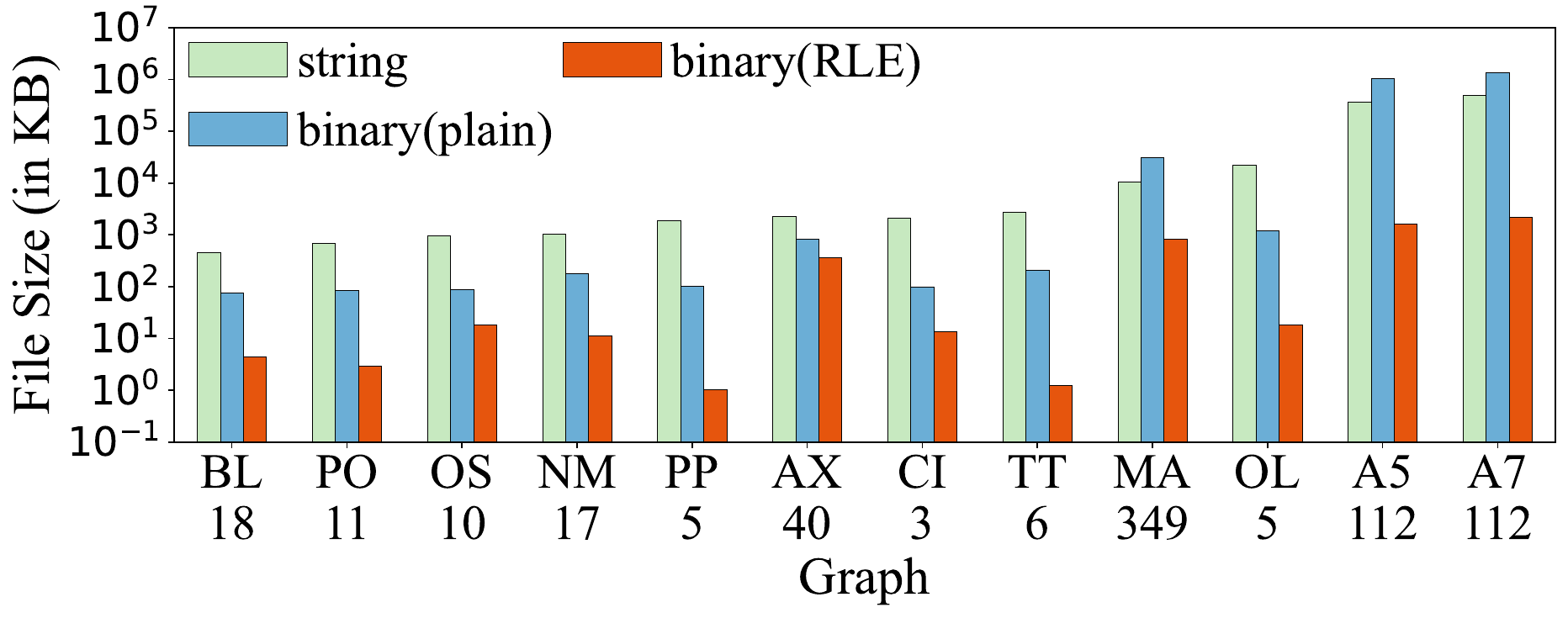}
        \vspace{-7mm}
        \caption{\revise{Storage consumption for labels.}}
        \label{fig:opt2-encoding}
    \end{subfigure}
    \hfill
    \begin{subfigure}{0.48\linewidth}
        \centering
        \includegraphics[width=\linewidth]{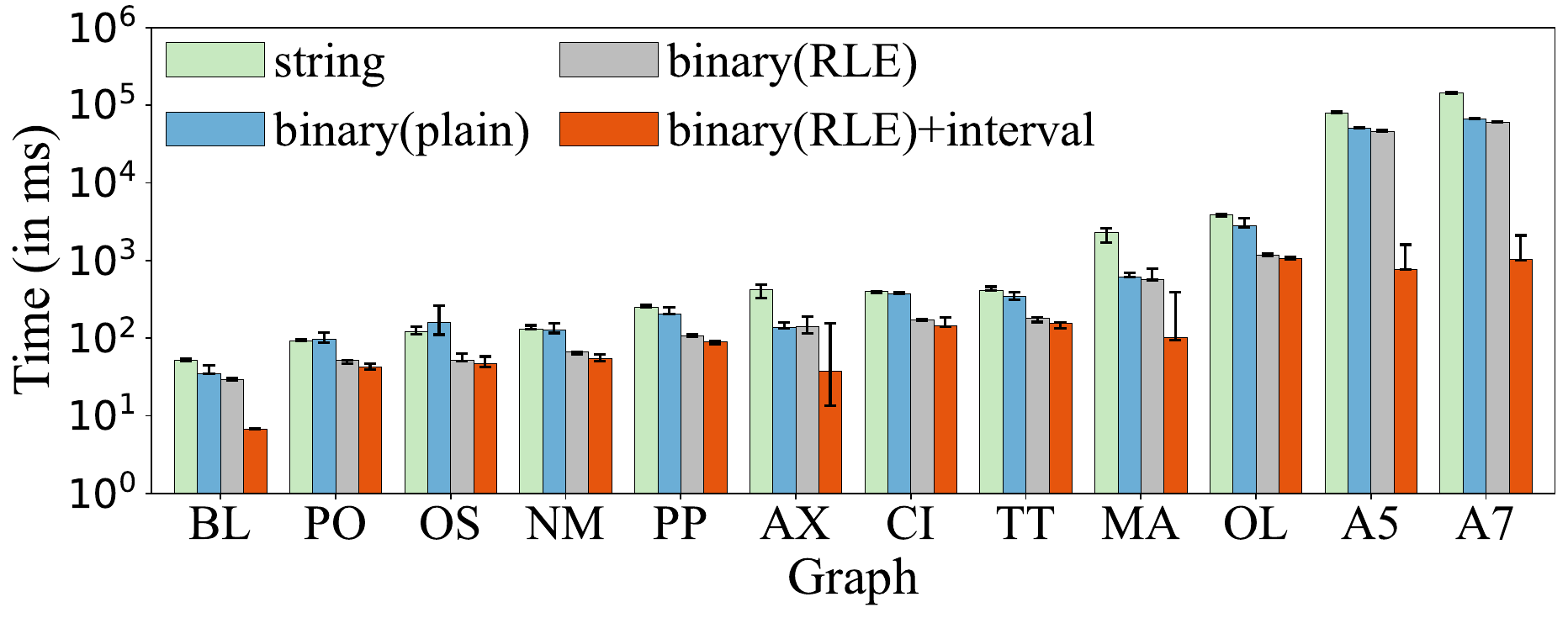}
        \vspace{-7mm}
        \caption{\revise{Simple condition filtering.}}
        \label{fig:opt2-simple}
    \end{subfigure}
    %\hfill

    \vspace{2mm}

    \begin{subfigure}{0.48\linewidth}
        \centering
        \includegraphics[width=\linewidth]{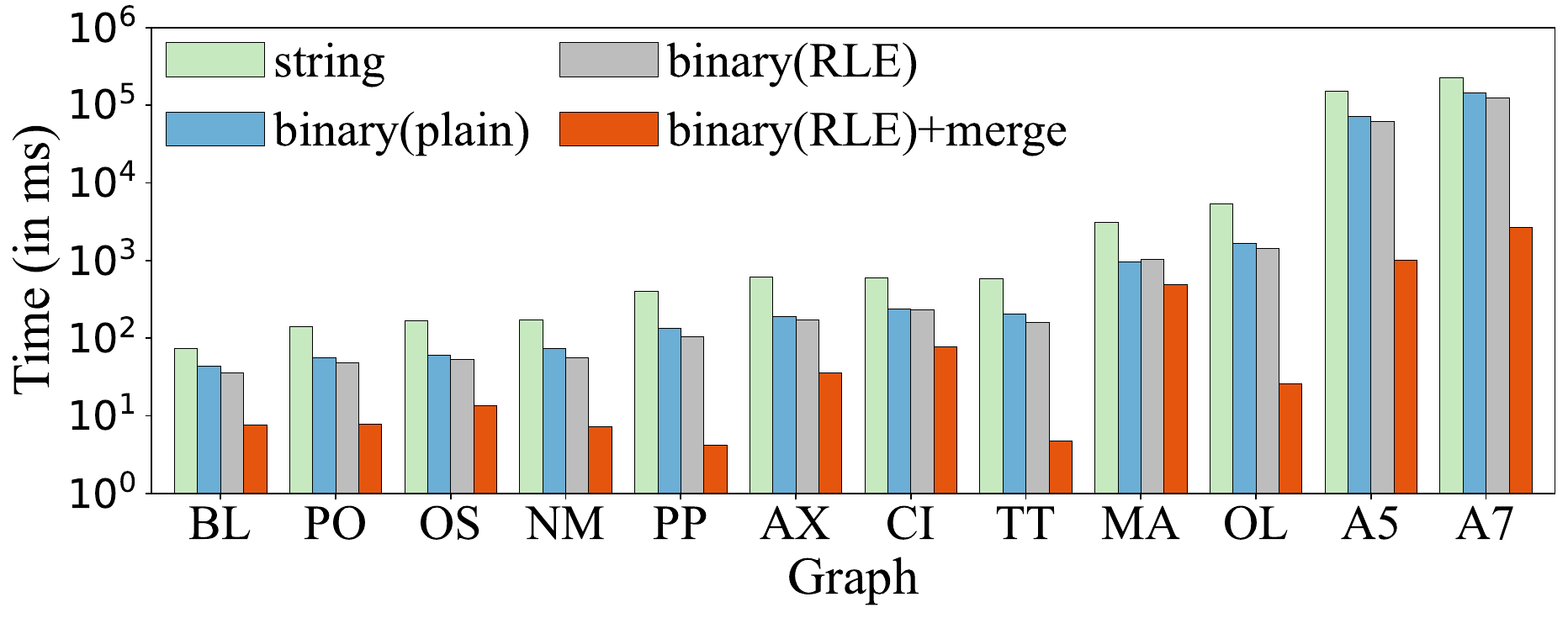}
        \vspace{-7mm}
        \caption{\revise{Complex condition filtering.}}
        \label{fig:opt2-complex}
    \end{subfigure}
    \hfill
    \begin{subfigure}{0.48\linewidth}
        \centering
        \includegraphics[width=0.98\linewidth]{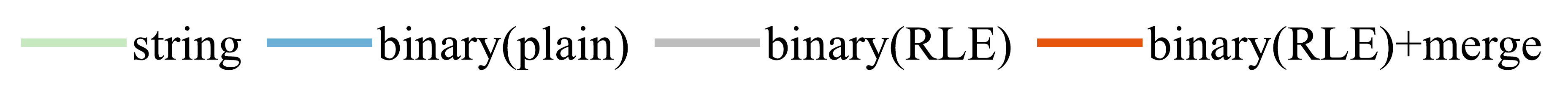}
        \begin{minipage}{0.49\linewidth}
            \centering
            \includegraphics[width=\linewidth]{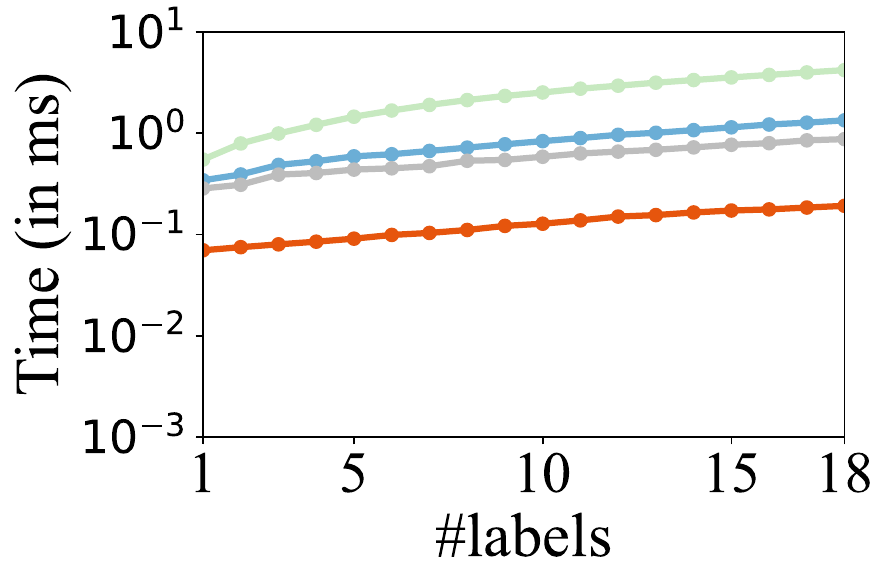}
        \end{minipage}
        \hfill
        \begin{minipage}{0.49\linewidth}
            \centering
            \includegraphics[width=\linewidth]{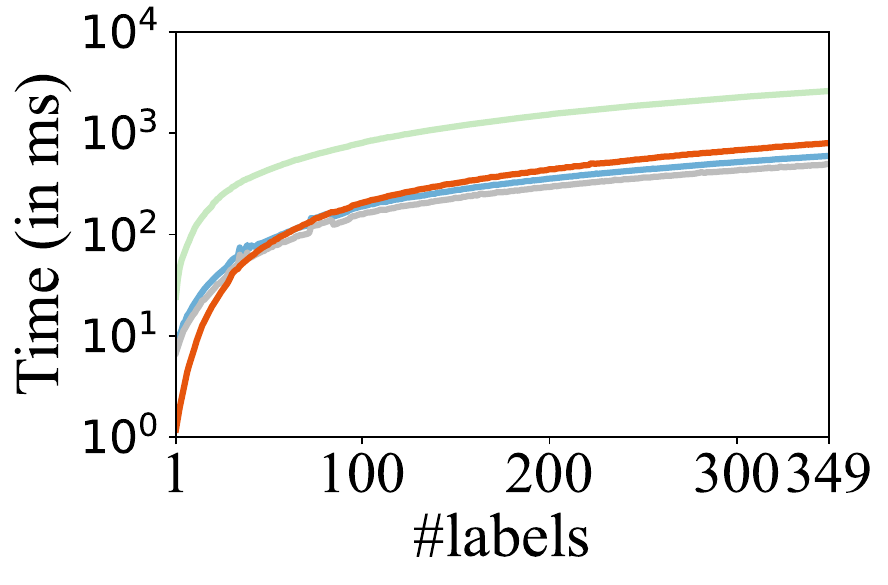}
        \end{minipage}
        \vspace{-3mm}
        \caption{Scaling the number of labels, tested on BL (left) and MA (right).}
        \label{fig:opt2-scale}
    \end{subfigure}

    \vspace{-1mm}
    \caption{Micro-benchmark of label filtering.}
    \vspace{-2mm}
\end{figure*}

%\subsubsection{Performance of Data Transformation}
\stitle{Performance of data transformation.}
Given that \gar is designed for storing LPGs in data lakes, the efficiency of converting original graph data into the \gar format is crucial. 
Graphs generally have significantly more edges than vertices and \gar employs CSR/CSC-like layouts requiring edge sorting. Thus, generating topological data becomes the most time-intensive part.
To assess the overhead, we analyze the time to convert four real-world graphs (U2, AR, WB, and U5), each initially in the form of Arrow Tables, a standardized in-memory format in big data systems. 
%Since graphs generally have significantly more edges than vertices and \gar employs CSR/CSC-like layouts requiring edge sorting, generating topological data becomes the most time-intensive part.

\revise{
Figure~\ref{fig:opt1-generation} illustrates the time breakdown. 
\revisenote{M3 R2.W2}
%which shows that we can generate topological data for over $3.9$ million edges per second. 
The process involves three steps: 1) sorting the edges, using Arrow's \emph{order\_by} operator, labeled as ``sort (Arrow)''; 2) generating vertex offsets, labeled as ``offset''; and 3) writing the sorted and offset data into \parquet~files with specific encoding, labeled as ``write''. 
%Generally, this process has a time complexity of $O(|E|log|E|)$ sequentially. 
%The WB graph takes less time than U5 due to the clustered edge distribution by source in its original data, while U5's original data is randomly distributed. 
The sorting step is most time-consuming, which has an average time complexity of $O(|E|log|E|)$ when executed sequentially, while the steps of generating offset and writing with encoding both have a lower time complexity of $O(|E|)$.
For further optimization, we leverage the parallel sorting algorithms provided by Intel(R) Threading Building Blocks~\cite{TBB}, which significantly reduces the sorting time, labeled as ``sort (TBB)''. 
By employing 24 threads in our test setup, the sorting time is only $8.9\%$ of the original, on average. 
Given this transformation is a one-time, offline operation that substantially reduces future data retrieval times, the associated overhead—which is within 1 minute for generating topological data for over 1 billion edges—is acceptable.\looseness=-1
%The most time-consuming step, sorting, offers room for further optimization through distributed frameworks like Spark.
}

\subsection{Micro-Benchmark of Label Filtering}
This section evaluates \gar's efficiency in storing and filtering vertex labels. We employ datasets from OGB~\cite{hu2020ogb}, Neo4j~\cite{neo4j_dataset} and Alibaba's synthetic data generator, which feature property graphs with multiple vertex labels, with the number of labels (ranging from $3$ to $349$) indicated under the graph name in Figure~\ref{fig:opt2-encoding}.

%\subsubsection{Storage Efficiency}
%\label{sec:opt2-encoding}
\stitle{Storage efficiency.}
We assess storage efficiency by measuring the size of encoded \parquet~files used for storing vertex labels. Two baseline methods serve for comparison: The first, termed ``string'', concatenates all labels of a vertex into a single string column using \emph{BYTE\_ARRAY} datatype and plain encoding. The second, named ``binary (plain)'', represents each label in a separate binary column using \emph{BOOLEAN} datatype and plain encoding. Our approach, denoted as ``binary (RLE)'', further optimizes this by utilizing RLE.

\revisenote{M1 R1.W1}
\revise{
%As shown in Figure~\ref{fig:opt2-encoding}, our RLE-based method substantially outperforms the baselines, requiring on average only $2.9\%$ and $10.1\%$ of the storage space compared to the ``string'' and ``binary (plain)'' methods, respectively. 
As shown in Figure~\ref{fig:opt2-encoding}, our RLE-based method substantially outperforms the baselines, requiring on average only $2.5\%$ and $8.4\%$ of the storage space compared to the ``string'' and ``binary (plain)'' methods, respectively. 
}
While \parquet~does support dictionary encoding that could potentially enhance the ``string'' baseline, we excluded this from our evaluation. The reason being that, despite some storage gains, dictionary encoding incurs a decoding slowdown of up to $10\times$ due to the extra overhead of storing the dictionary, especially when label numbers are high. 
Our RLE-based strategy strikes a balance between storage efficiency and decoding performance, as demonstrated in the subsequent experiments.

%\subsubsection{Performance of Simple Condition Filtering}
\stitle{Performance of simple condition filtering.}
Recognizing that filtering based on simple conditions represents the cornerstone operation in graph query languages, we prioritize evaluating this operation.
For each graph, we perform experiments where we consider each label individually as the target label for filtering, and determine the vertices with that label.
For accuracy, each experiment is repeated $100$ times and the total execution time is reported.

Figure~\ref{fig:opt2-simple} illustrates the results, demonstrating that \gar's method  significantly improves the performance of label filtering based on simple conditions.
The most straightforward approach, ``string'', which involves decoding the string of labels and conducting matching for each vertex, is the slowest.
The ``binary (plain)'' method separates labels into individual columns and utilizes a binary representation, while the  `binary (RLE)'' method further optimizes the encoding by using RLE.
However, both of these methods still require evaluating each vertex.
In contrast, our method of ``binary (RLE) + interval'', simply selects all satisfied intervals.

In Figure~\ref{fig:opt2-simple}, for each graph, we report the middle value of the execution time among filtering each label as the height of the bar, with the error bar representing the range of execution time. 
Our method may have a large range on some graphs (AX, MA) due to the varying encoding efficiency (i.e., the number of intervals generated) for different columns.
However, since the number of intervals is not larger than the number of vertices in any case, our method consistently outperforms the baselines.
\revisenote{M1 R1.W1}
\revise{
%On average, it achieves a speedup of $6.0\times$ over the ``string'' method, $3.3\times$ over the ``binary (plain)'' method, and $2.3\times$ over the ``binary (RLE)'' method.
On average, it achieves a speedup of $14.8\times$ over the ``string'' method, $8.9\times$ over the ``binary (plain)'' method, and $7.4\times$ over the ``binary (RLE)'' method.
}

%\subsubsection{Performance of Complex Condition Filtering}
\stitle{Performance of complex condition filtering.}
We also assess the performance of label filtering based on complex conditions using the same graphs as mentioned above. 
For graphs obtained from Neo4j, we first refer to the provided documentation to identify a filtering operation that involves two labels.
If not provided, we create a condition by combining two related labels using either the logical \emph{AND} operator (if there are vertices satisfying the condition) or \emph{OR} (otherwise) to reflect real-world semantics.
For other graphs, we combine the first two labels by \emph{OR} as the filtering condition.
%For graphs obtained from OGB, we utilize the first two labels and combine them using \emph{OR} as the filtering condition.

Figure~\ref{fig:opt2-complex} presents the performance of different methods, measured as the total execution time of $100$ runs.
The results demonstrate that \gar~performs the best for all test cases. 
Further analysis reveals that the performance improvement is attributed to the binary representation (as seen in the comparison between ``binary (plain)'' and ``string''), and utilization of RLE (as seen in the comparison between ``binary (RLE)'' and ``binary (plain)'').
\revisenote{M1 R1.W1}
\revise{
However, the merge-based decoding method yields the largest gain, where ``binary (RLE) + merge'' outperforms the ``binary (RLE)'' method by up to $60.5\times$.
}

%\subsubsection{Scale Up the Number of Labels}
\stitle{Scale up the number of labels.}
Figure~\ref{fig:opt2-scale} illustrates the average execution time of filtering conditions with varying numbers of labels, focusing on BL and MA, which are selected from different datasets (Neo4j and OGB) and have a relatively large number of labels.
%We test the filtering with $i$ labels by combining the first $i$ labels in a graph using the \emph{OR} operator as the condition, to ensure that there are vertices satisfying the condition.
We test the filtering with $i$ labels by combining the first $i$ labels through \emph{OR} as the condition. %, to ensure that there are vertices satisfying the condition.
%To reduce experimental noise, we conduct the experiments for multiple times and report the average time.

As shown in the figure, \gar~consistently outperforms others on BL.
While on MA, it performs best when the number of involved labels is no more than $40$.
As the number of labels continuously increases, it performs worse than the baseline ``binary (RLE)'', which is due to the number of merged intervals also increasing.
In the worst case, the UDF is called for each vertex, means any two consecutive vertices have different labels.
Considering the overhead of merging intervals from different columns, our method may perform worse than directly evaluating the UDF for each vertex.
Fortunately, our investigation of real-world workloads reveals that the number of filtered labels in user-written queries is often limited. For instance, in the Neo4j documentation examples ~\cite{neo4j_dataset}, the filtering involves at most $5$ labels, suggesting our method is highly promising.
%. Consequently, the performance of our method remains highly promising.

\subsection{Storage Media}
We assess the efficiency of \gar across various storage media: local in-memory \emph{tmpfs}, \emph{ESSD} (an Alibaba Cloud virtualized elastic block device), and S3-like Object Store Service (\emph{OSS}). The graph used is SF100, with specifically focus on the \emph{comment} vertex type and \emph{comment\_hasTag\_tag} edge type.
Table~\ref{tab:performance_comparison} encapsulates the efficiency of \gar across different storage media. These results demonstrates that \gar is not only efficient but also robust, delivering consistently high performance, with speedups of $88\times$ to $154\times$ for neighbor retrieval and $2.7\times$ to $11\times$ for label filtering.

\begin{table}[t]
    %\vspace{-3mm}
    \centering
    \caption{Performance comparison across storage media.}
    \vspace{-2mm}
   \fontsize{8}{9}\selectfont
    \label{tab:performance_comparison}
    \begin{tabular}{|c|c|c|c|c|}
        \hline
        & \multicolumn{2}{c|}{Neighbor Retrieval (s)} & \multicolumn{2}{c|}{Label Filtering (s)} \\
        \cline{2-5}
        Storage & Plain & \gar & String & \gar \\
        \hline
        tmpfs & 6.446 & 0.053 & 3.984 & 1.489  \\
        ESSD & 16.41 & 0.106 & 19.06 & 1.746 \\
        OSS & 189.4 & 2.145 & 252.8 & 26.22  \\
        \hline
    \end{tabular}
    \vspace{-4mm}
\end{table}

\subsection{End-to-end Graph Query Workloads}
To demonstrate the practicality of \gar~in real-world scenarios, we conduct a performance evaluation using end-to-end workloads from the LDBC SNB benchmark~\cite{LDBC, ldbc-bi}, which is widely used in the graph processing community. 
Although the benchmark specifies vertex/edge types, it does not explicitly define the labels.
However, we are able to identify certain vertex types that are \emph{static} (e.g., \emph{tagclass} and \emph{place}), which have a fixed and very small vertex set size that does not scale with the graph size. 
On the other hand, vertex types like \emph{comment} and \emph{person} are considered \emph{dynamic}.
Based on this observation, we can treat information related to \emph{static} types as labels for \emph{dynamic} types in \gar, for example, all tag classes of a comment are attached as labels for the corresponding \emph{comment} vertex.
Similar strategies are also adopted by graph databases~\cite{list-based} to optimize data access performance.

For evaluation, graphs at different scales (listed in Table~\ref{tab:datasets}) are generated using the LDBC SNB data generator. %, as \parquet~files.
These graphs are then converted into \gar~format, with the vertex labels attached as described above.
Upon investigating the benchmark, including $7$ short and $14$ complex interactive queries, as well as $20$ business intelligence queries,
we find that neighbor retrieval is frequently encountered, involved in approximately $90\%$ of the queries.
Considering the aforementioned label organization, label filtering is also common, involved in approximately $50\%$ queries.

%\subsubsection{Query Implementations}

%\stitle{Queries.} 
\stitle{Query implementations.}
The evaluation focuses on three representative queries, with the required parameters set according to the reference implementations~\cite{ldbc_implementation, bi_implementation}.
%The first query, \textbf{IS-3} (\emph{interactive-short-3}), aims to find all the friends of a given person and return their information.
\textbf{IS-3} (\emph{interactive-short-3}) aims to find all the friends of a given person and return their information.
It exemplifies the common pattern of querying neighboring vertices and retrieving associated properties.
%The second query, \textbf{IC-8} (\emph{interactive-complex-8}), is more complex as it involves traversing multiple hops from the starting vertex.
\textbf{IC-8} (\emph{interactive-complex-8}) is more complex as it involves traversing multiple hops from the starting vertex.
Lastly, the \textbf{BI-2} (\emph{business-intelligence-2}) query involves finding and counting the messages associated with tags within a specific tag class, thus requiring vertex filtering by labels.

%\stitle{\gar.} 
We develop hand-written implementations for each query based on \gar, which utilize the data organization and specifically prioritize two essential operations: neighbor retrieval and label filtering.
Our implementation adheres to the official reference implementations~\cite{ldbc_implementation, bi_implementation} to ensure equivalence to the original queries.

%\stitle{Acero.} 
We then implement these queries in Acero~\cite{Acero}, which is a powerful C++ library integrated into Apache Arrow for analyzing large streams of data.
It offers a comprehensive set of operators such as \emph{scan}, \emph{filter}, \emph{project}, \emph{aggregate}, and \emph{join}, among others.
Moreover, Acero supports taking \parquet~as the data source and enables the pushdown of predicates, making it a strong baseline for comparison with \gar.
Despite our best efforts to optimize it, we do not perform data re-organization or utilize \gar's encoding/decoding optimizations for this implementation based on Acero.

\revisenote{M4 R2.W3 R2.D2}
\revise{
We also include two additional baselines: Apache Pinot~\cite{pinot}, a real-time OLAP datastore used by LinkedIn for processing and querying large social networks, and Neo4j~\cite{neo4j}, a main graph database utilizing the Cypher query language.
%and Apache TinkerPop~\cite{tinkerpop}, a framework supporting various storage backends and providing a common Gremlin interface. 
While both are widely-used, they are not natively designed for data lakes and require an Extract-Transform-Load (ETL) process for integration.
}

\begin{table}[t]
    \centering
    \caption{\revise{Query execution times (in seconds), with the format of \textit{Pinot (P), Neo4j (N), Acero (A), \gar (G)}. ``OM'' denotes failed execution due to out-of-memory errors.}}
    \vspace{-2mm}
    \label{tab:workload}
    \setlength{\tabcolsep}{2pt}
    \fontsize{8}{9}\selectfont
    %\begin{tabular}{|l|rrrr|rrrr|rrrr|}
    \begin{tabular}{|c|cccc|cccc|cccc|}
    \hline
    & \multicolumn{4}{c|}{SF30} & \multicolumn{4}{c|}{SF100} & \multicolumn{4}{c|}{SF300}\\
    & P & N & A & G & P & N & A & G & P & N & A & G\\
    \hline
    ETL & 6024 & 390 & --- & --- & 17726 & 2094 & --- & --- & OM & 9122 & --- & --- \\
    \hline
    IS-3 & 1.00 & 0.30 & 0.16 & \textbf{0.01} & 6.59 & 2.09 & 0.48 & \textbf{0.01} & OM & 4.12 & 1.39 & \textbf{0.03} \\
    IC-8 & 1.35 & \textbf{0.37} & 72.2 & 3.36 & 8.43 & \textbf{1.26} & 246 & 6.56 & OM & \textbf{2.98} & 894 & 23.3 \\
    BI-2 &  125 & 45.0 & 67.7 & \textbf{4.30} & 3884 & 1101 & 232 & \textbf{16.3} & OM & 6636 & 756 & \textbf{50.0} \\
    \hline
\end{tabular}
\vspace{-4mm}
\end{table}

\eat{
\begin{table}[t]
    \centering
    \caption{Execution time for IS-3, IC-8 and BI-2 (in seconds).}
    \vspace{-2mm}
    \label{tab:workload}
    \fontsize{8}{9}\selectfont
    \begin{tabular}{|l|rr|rr|rr|}
    \hline
    Graph & \multicolumn{2}{c|}{SF30} & \multicolumn{2}{c|}{SF100} & \multicolumn{2}{c|}{SF300}\\
    Impl. & Acero & \gar~& Acero & \gar~& Acero & \gar\\
    \hline
    IS-3 & 0.156 & 0.005 & 0.475 & 0.010 & 1.390 & 0.029 \\ 
    IC-8 & 72.22 & 3.362 & 245.5 & 6.563 & 894.4 & 23.29 \\
    BI-2 & 67.74 & 4.295 & 231.6 & 16.28 & 755.6 & 50.04 \\
    \hline
\end{tabular}
\vspace{-8mm}
\end{table}
}

%\subsubsection{Performance Comparison}
\stitle{Performance comparison.}
Table~\ref{tab:workload} presents a comparison of end-to-end performance, clearly demonstrating that the implementation based on \gar~significantly outperforms Acero, achieving an average speedup of $29.5\times$.
%Upon closer examination of the results, it becomes apparent that the performance improvement can be attributed to the following factors: 
A closer analysis of the results reveals that the performance gains stem from the following factors:
1) data layout design and encoding/decoding optimizations we proposed, to enable efficient neighbor retrieval (IS-3, IC-8, BI-2) and label filtering (BI-2), as demonstrated in micro-benchmarks; 2) bitmap generation during the two critical operations, which can be utilized in subsequent selection steps (IS-3, IC-8, BI-2).

%\revisenote{M4 R2.W3 R2.D2}
\revise{
As for Pinot and Neo4j, their end-to-end performance is often dominated by extensive ETL processes, in the context of data lakes, as the results show.
\gar performs best on IS-3, which is a single-hop query, and BI-2, where \gar utilizes label filtering for the early elimination of irrelevant data.
While on IC-8, \gar is outperformed by Neo4j due to the query involving traversing multiple hops, which results in a significant volume of data loading for both Acero and \gar.
Nevertheless, \gar not only offers efficient query performance but also eliminates the ETL overhead, potentially avoids out-of-memory errors that may occur.
Thus, \gar provides a more practical solution for data lake scenarios.
}

%Taking IS-3 as an example, the optimization of neighbor retrieval achieves an average speedup of $571.4\times$ compared to the baseline utilizing \parquet's filter pushdown. 
%The subsequent selection based on bitmaps provides an average speedup of $2.7\times$.
%Other graph-agnostic steps like \emph{order\_by} show similar performance in \gar~and Acero due to closely matched implementations.

\vspace{-3mm}
\subsection{Integration with Graph Processing Systems}
One of the advantages of \gar~is its compatibility with existing graph processing systems, enabling seamless integration.
To demonstrate this, we have successfully integrated \gar~into GraphScope~\cite{graphscope} as the archive format for persistent storage. 
In addition, it is also utilized as an accessible storage backend for executing infrequent queries in an out-of-core manner.
GraphScope is a distributed system for addressing a wide array of graph computing need, which has achieved significant technological advancements and gained widespread adoption across various industries.

\begin{figure}[tbp]
    \centering
    \begin{subfigure}{0.48\linewidth}
        \centering
        \includegraphics[width=\linewidth]{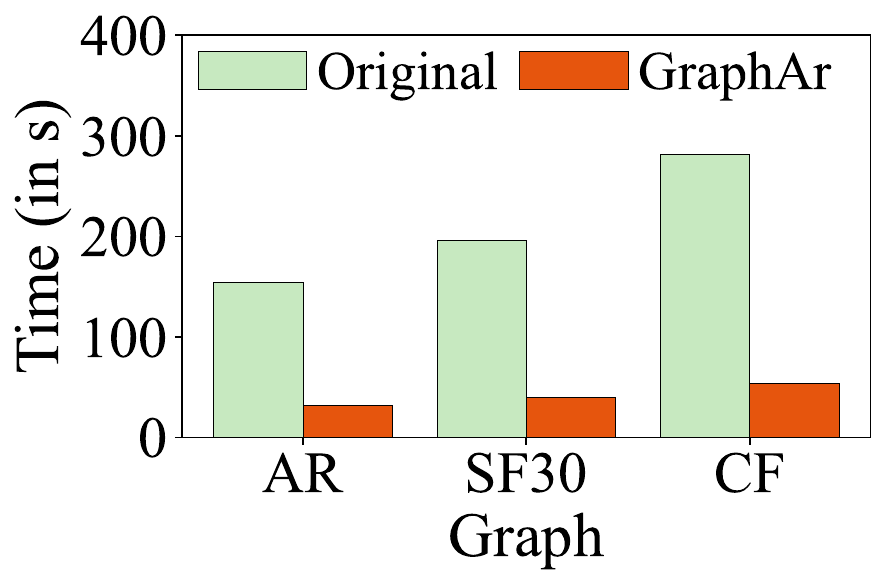}
        \vspace{-7mm}
        \caption{Graph loading time.}
        \label{fig:gs-load}
    \end{subfigure}
    \hfill
    \begin{subfigure}{0.48\linewidth}
        \centering
        \includegraphics[width=\linewidth]{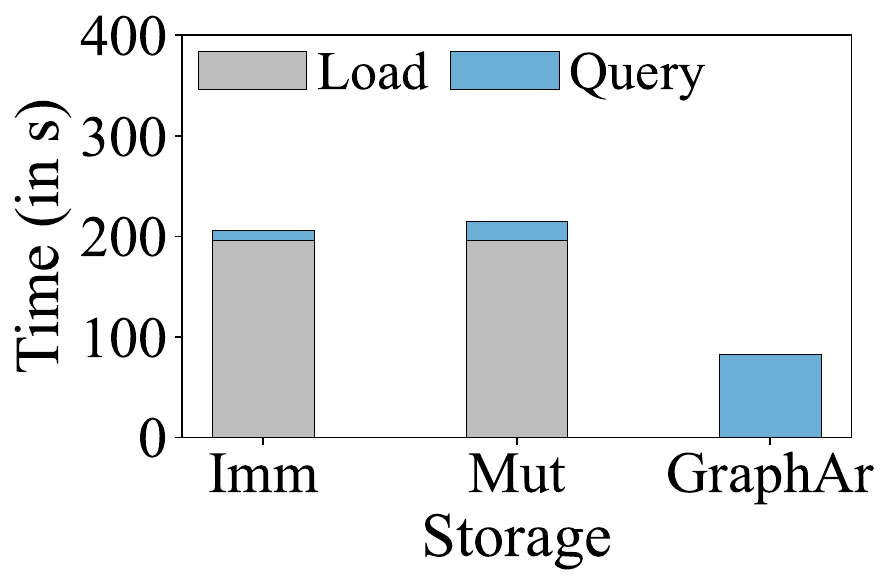}
        \vspace{-7mm}
        \caption{BI queries execution time.}
        \label{fig:gs-query}
    \end{subfigure}
    \vspace{-1mm}
    \caption{GraphScope's performance w/ and w/o \gar.}
    \vspace{-5mm}
\end{figure}

\stitle{Serve as the archive format.} We first compare the performance of building graphs in GraphScope using 8 nodes, from external storages in \gar format against the baseline, where the datasets are in CSV format, sourced directly from the data providers.
The findings illustrated in Figure~\ref{fig:gs-load} indicate that \gar significantly outperforms the baseline, achieving an average speedup of $4.9\times$.
This improvement can be attributed to \gar's efficient encoding strategies that reduce the data volume to be loaded, as well as its optimized data organization and layout, which facilitate a faster in-memory graph construction within GraphScope.

\stitle{Serve as a storage backend.} Leveraging the capabilities for graph-related querying, the graph query engine within GraphScope can execute queries directly on the \gar data in an out-of-core manner.
%This approach is especially beneficial for data that is accessed infrequently, or when the entire graph is too large to fit into memory, as it obviates the necessity to load the entire graph and construct an in-memory representation prior to processing.
We evaluate the performance of GraphScope with \gar as the storage backend, and compare the average querying time of BI queries on SF30 with two baseline scenarios wherein GraphScope relies on its native in-memory storage options, specifically the immutable (``Imm'') and mutable (``Mut'') variants.
Figure~\ref{fig:gs-query} demonstrates that although the querying time with \gar exceeds that of the in-memory storages, attributable to intrinsic I/O overhead, it significantly surpasses the process of loading and then executing the query, by $2.4\times$ and $2.5\times$, respectively.
This indicates that \gar is a viable option for executing infrequent queries.
% in an out-of-core manner.

\stitle{\revise{Application scenarios.}} 
\revisenote{R2.D3}
\revise{This integration demonstrates the potential benefits of \gar in improving the efficiency of graph processing systems. In summary, its application scenarios include: 1) \emph{Data loading:} \gar can significantly reduce graph loading times, making it an ideal choice for external storage formats; 2) \emph{Out-of-core queries:} \gar can serve as a storage backend for executing graph queries in an out-of-core manner. It is particularly beneficial for infrequent queries that access only a portion of the graph data, eliminating the need for full in-memory graph representation. It also enables querying graphs that exceed the capacity of available memory.
In other scenarios, such as: 1) real-time queries requiring low latency and frequent execution, or 2) graph analytics algorithms involving iterative computations across the entire graph (e.g., PageRank), \gar might not be the optimal direct storage solution and in-memory storage options are more suitable.}

\sstab
{\bf Summary.}
In the comprehensive evaluation, \gar~has been demonstrated to be a highly effective storage scheme for LPGs in data lakes. The key takeaways are:
\begin{itemize}
    %\vspace{-1mm}
    \setlength{\itemsep}{-1pt}
    \item \textbf{Storage efficiency:} \gar~remarkably reduces storage requirements, using only \(29.2\%\) of the storage compared to baseline methods for storing topology, and as low as \(2.5\%\) for label storage on average.
    %\item \textbf{Storage efficiency:} \gar~remarkably reduces storage requirements, using only \(27.3\%\) of the storage compared to baseline methods for storing topology, and as low as \(2.9\%\) for label storage on average.
    \item \textbf{Query performance:} \gar~significantly outpaces the baselines in retrieval time, achieving an average speedup of \(4452\times\) for neighbor retrieval, and an average speedup of \(14.8\times\) for simple label filtering, as observed in micro-benchmarks on the \emph{ESSD} storage.
    %\vspace{-1.5ex} 
    \item \textbf{Storage media:} Evaluations indicate seamless compatibility across various storage layers like \emph{tmpfs}, \emph{ESSD}, and \emph{oss}, all achieving high speedup of $88\times$ to $154\times$ for neighbor retrieval and $2.7\times$ to $11\times$ for label filtering.
    %\vspace{-4ex} 
    \item \textbf{Real-world relevance:} In end-to-end workloads using the LDBC SNB benchmark, \gar~shows an average speedup of \(29.5\times\) over the Acero baseline, substantiating its practical utility in real-world scenarios.
    \item \textbf{Compatibility:} \gar~is seamlessly integrated into a widely-used graph processing system GraphScope, enhancing its graph loading efficiency by \(4.9\times\), and accelerating its infrequent query execution with a speedup of \(2.4\times\).
    %\vspace{-2mm}
\end{itemize}
Collectively, these results validate \gar~as a robust, storage-efficient, and high-performance solution for both academic research and industrial applications.
\section{Related Work}
\label{sec:related}
%\lx{TODO: add related work on the following topics: (1) review existing archiving format for relational data; (2) other formats besides ORC/Parquet for graph or RDF, like hdf5; (3) data lake and delta lake; (4) pushdown optimizations and decoding techniques based on instructions.}
% \lx{TODO: refine this}
\stitle{File formats in data lakes.} The data lake ecosystem encompasses various common file formats, including CSV, JSON,  Protocol Buffers, HDF5~\cite{hdf5}, AVRO~\cite{avro}, ORC~\cite{orc}, and Parquet~\cite{parquet}.
%While these formats support various optimizations that benefit both tables and graphs, they do not inherently cater to some unique needs of LPGs, thus fall short in comprehensively representing LPG semantics and supporting graph-specific operations.
While these formats support various optimizations that benefit both tables and graphs, they fall short in comprehensively representing LPG semantics and supporting graph-specific operations.

\revisenote{M4 R2.W3 R2.D2}
\revise{
\stitle{Data management in data lakes.} The popularity of data lakes has led to efforts aimed at enhancing their architecture and data management~\cite{data_lake_management, photon, puppygraph, Metadata, fishing}.
%These endeavors primarily focus on managing existing data within data lakes and are distinct from  \gar. 
%\gar, on the other hand, can be considered as a new storage format with unique features tailored for LPGs.
%It can be leveraged by these works to further extend the utility and capabilities of data lakes.
As for LPG management in data lakes, LinkedIn uses Apache Gobblin~\cite{gobblin} for data ingestion and employs Apache Spark~\cite{zaharia2016apache} and Apache Pinot~\cite{pinot} for processing large graph datasets representing the social network. Graph-specific querying frameworks like Neo4j~\cite{neo4j} and Apache TinkerPop~\cite{tinkerpop} are also widely-used, and they integrate with data lakes via ETL processes. 
These endeavors primarily focus on managing existing data within data lakes and are distinct from  \gar. 
\gar, on the other hand, can be considered as a new storage format with unique features tailored for LPGs.
It can be leveraged by these works to further extend the utility and capabilities of data lakes.
}

\stitle{Graph file formats.} Certain formats are designed for graph~\cite{graphml, Himsolt2010GMLAP, gexf, batagelj2001pajek} and RDF (Resource Description Framework) data~\cite{rdf, hdt}. 
However, their primary focus is to describe or exchange data in a standardized manner, e.g., utilizing XML, and are not optimized for storage and retrieval purposes.
The lack of encoding, compression and push-down optimizations can lead to far inferior performance, making them less suitable for managing LPGs in data lakes.

\stitle{Graph-related databases.} %There are some graph databases~\cite{neo4j, TigerGraph, janusgraph,feng2023kuzu} that have been developed to store and manage graph data. 
%There are also efforts focus on optimizing graph-related queries~\cite{nguyen2015join, zhao2017all,mhedhbi2022modern, raasveldt2019duckdb}, such as by mapping graph queries to a sequence of join operations.
Some databases~\cite{neo4j, TigerGraph, janusgraph,feng2023kuzu} are designed to store and manage graph data. 
There are also efforts focus on optimizing graph-related queries~\cite{nguyen2015join, zhao2017all,mhedhbi2022modern, raasveldt2019duckdb}.
While they offer various graph-related features, they primarily focus on in-memory mutable data management, operating at a higher level compared to \gar. \gar, with its format compatible with the LPG model, can be utilized as an archival format for graph databases.
%While they offer various features that are tailored for graphs, they primarily focus on in-memory mutable data management, operating at a higher level compared to \gar. \gar, with its format compatible with the LPG model, can be utilized as an archival format for graph databases.

\stitle{Operation pushdown.} Some previous works~\cite{li2023selection, Vectorizing2013, simd4, FlexPushdownDB, PushdownDB, AWS_S3_Select} aim to develop high-performance operators on storage formats of either column-oriented or row-oriented.
%These works and \gar~share the same goal of improving pushdown operators and making the utilization of available CPU instructions like SIMD and BMI.
These works and \gar~share the same goal of improving pushdown operators and leveraging CPU instructions like SIMD and BMI.
%However, it is important to note that these works mainly focus on operations related to relational data, such as scan, select, and filter based on properties. 
However, these works mainly focus on operations related to relational data, such as scan, select, and filter based on properties. 
In contrast, \gar~specifically focuses on two graph-specific operations.

\section{Conclusion}
In conclusion, this paper introduces \gar~as an efficient and specialized storage scheme for graph data in data lakes.
\gar~focuses on preserving LPG semantics and supporting graph-specific operations, resulting in notable performance improvements in both storage and query efficiency over existing formats designed for relational tables.
The evaluation results validate the effectiveness of \gar~and highlight its potential as a crucial component in data lake architectures.

%\clearpage

\bibliographystyle{ACM-Reference-Format}
\bibliography{paper}

\end{document}
\endinput